\documentclass[aps,prb,onecolumn]{revtex4-2}

\usepackage{amsmath}
\usepackage{amssymb}
\usepackage{graphicx}
\usepackage{bm}
\usepackage{ragged2e}
\usepackage{braket}
\usepackage{comment}
\usepackage{subcaption}
\usepackage[colorlinks=true,linkcolor=blue,citecolor=blue,urlcolor=blue]{hyperref}

\begin{document}

\title{Analytical classification of Majorana zero-mode spatial profiles in extended Kitaev chains: probability maxima can shift inward}

\author{Vijay Pathak}
\email{vijaypathak.iisertvm@gmail.com}

\author{Vaishnav Mallya}
\author{Sujit Sarkar}

\affiliation{Theoretical Sciences Department, Poornaprajna Institute of Scientific Research, Bengaluru, India 562 164}

	%\keywords{Kitaev chain, Majorana zero modes, topological superconductivity, extended interactions, recursion relation, bulk-boundary correspondence}
	
	\begin{abstract}
    Topological phases in one-dimensional superconducting systems are commonly characterized by symmetry-protected invariants. These invariants determine the number of Majorana zero-energy boundary modes but do not specify their corresponding spatial structure. In this work, we present an analytical study of Majorana zero modes (MZMs) in an extended Kitaev chain with nearest- and next-nearest-neighbor couplings. By expressing the Hamiltonian in the Majorana basis, we derive a recursion relation whose characteristic roots completely determine the spatial structure of the zero modes and yield closed-form expressions for their amplitudes. We show that, even within a single topological phase, the MZMs can exhibit qualitatively distinct decay behaviors - monotonic decay, oscillatory decay, and perfectly localized states. Remarkably, boundary-origin MZMs need not have their maximum probability at the edge of the chain. They can instead exhibit maxima at interior lattice sites with an exponentially decaying envelope from either side of the maxima. Furthermore, the characteristic roots determine the length scale required for finite chains to reproduce the semi-infinite MZM structure, providing a direct link between Hamiltonian parameters, finite-size effects, and experimentally observable spatial profiles. 
	\end{abstract}
	
	\maketitle
    
\section{Introduction}
Majorana zero modes (MZMs)~\cite{Moessner_Moore_2021, stanescu2024introduction, lutchyn2018majorana, fu2008superconducting} have attracted significant attention in recent years because of their potential applications in topological quantum computation~\cite{KITAEV20032, aasen2016milestones, sarma2015majorana, RevModPhys.80.1083, PhysRevB.88.064515, Freedman2003}. In a seminal work, Kitaev showed that a one-dimensional p-wave superconducting chain can host zero-energy Majorana modes localized at its ends~\cite{Kitaev2001}. These modes correspond to quasiparticle operators that are self-conjugate and satisfy Majorana fermion statistics. Because the information encoded in spatially separated Majorana modes is non-local, it is expected to be protected against certain local perturbations and decoherence. A growing number of experiments have reported signatures consistent with Majorana-like excitations in engineered superconducting systems~\cite{law2009majorana, dvir2023realization, bordin2024crossed, Beenakker2013, jack2019observation, tenHaaf2025, PhysRevB.111.235409}. However, the unambiguous realization of topologically protected Majorana modes remains challenging. In current experiments, the superconducting chains are typically short, which can lead to hybridization between modes localized at opposite ends of the chain, resulting in small but finite energy splittings~\cite{leijnse2012parity, bordin2025enhanced, tenHaaf2025, leumer2020exact, PhysRevB.111.115419}. Consequently, zero-energy features observed in such systems may originate from finite-size effects rather than from perfectly isolated Majorana-bound states. Determining how the spatial structure and localization of zero modes depend on system parameters and system size, and therefore remains an important open problem to address~\cite{svensson2024quantum, PhysRevB.111.115419}.

A prominent model system for studying MZMs is the one-dimensional Kitaev chain~\cite{Kitaev2001}. A key concept underlying the appearance of MZMs in one-dimensional systems is the presence of symmetry-dependent topological phases and their corresponding invariants. In the Kitaev chain and related models, these phases are characterized by an integer winding number, and a topological phase transition occurs when the excitation gap closes and reopens. Through bulk-boundary correspondence, the value of this invariant determines the number of boundary MZMs present in a given phase~\cite{Niu_2012, rahul2021majorana}. As long as the bulk gap remains open, continuous variations of system parameters within a phase do not change the invariant and therefore do not alter the number of boundary modes. While the topological invariant fixes the number of boundary modes, it does not determine their corresponding spatial structure or localization properties. The spatial profile of the MZMs is particularly important in experimental realizations, where measurements probe the local probability distribution rather than the topological invariant itself. Consequently, different parameter choices within the same topological phase can yield qualitatively distinct spatial structures of MZMs. 

While the original model contains only nearest-neighbor hopping and pairing terms, realistic systems such as the Kitaev chain can include long-range interactions. Such an extension of the Kitaev model provides a richer platform for investigating the behavior of MZMs, and can support additional topological phases characterized by higher winding numbers~\cite{Alecce_2017, kopp2005criticality, rahul2021majorana, svensson2024quantum, mahyaeh2018zero, sticlet2013, Niu_2012}. Such models have been studied in a variety of contexts, including generalized Kitaev chains and related spin systems obtained through the Jordan-Wigner transformation~\cite{kopp2005criticality, rahul2021majorana, svensson2024quantum, mahyaeh2018zero, sticlet2013, Niu_2012, PhysRevB.92.115115, fraxanet2021topological, ares2018entanglement, baghran2024competition, vodola2014kitaev, kumar2022physics, sarkar2020study, sarkar2018quantization}. In the present work, we focus on a particular subclass of the generalized Kitaev model in which the hopping and pairing amplitudes are equal. This choice of parameters significantly simplifies the analysis and allows the Hamiltonian to be expressed in a Majorana representation, yielding a quadratic recursion relation for the MZM amplitudes. For more general parameter choices, the recursion relation becomes a higher-order polynomial equation whose analytic solution is generally intractable, and one must typically rely on numerically diagonalizing finite systems.

Despite these developments, a general analytical framework that directly characterizes the spatial structure of MZMs in extended Kitaev chains remains lacking. Existing studies have largely focused on numerical analyses of finite systems or on specific limiting cases, making it difficult to distinguish intrinsic properties of the zero modes from the finite-size effects. This motivates the search for an analytical solution that directly reveals the ideal spatial structure of the MZMs, independent of system size. This limit is particularly important because experimentally accessible systems are finite, where boundary-induced hybridization and exponentially small energy splittings can obscure intrinsic MZM properties. Without analytical guidance, it is often difficult to determine whether features such as imperfect localization or shifts in the position of the probability maximum arise from the underlying Hamiltonian or from insufficient system size.

To address this gap, we analyze the recursion relations governing MZMs in different parameter regimes and obtain their analytical solutions. This identifies the intrinsic zero-energy structure in the semi-infinite limit, determined solely by the underlying Hamiltonian, in which boundary-induced hybridization is absent, and the ideal features of the modes within a given topological phase can be isolated. We find that MZMs can exhibit distinct spatial behaviors depending on system parameters, including monotonic decay, oscillatory decay, and perfectly localized profiles. Moreover, these zero modes naturally satisfy Majorana properties and hence correspond to MZMs. These behaviors are determined by the structure of the characteristic roots of the recursion relation and their dependence on the Hamiltonian parameters. A direct consequence is that the spatial distribution of the MZMs need not be maximized at the boundaries. Boundary-origin MZMs can also attain their maximum probability at interior lattice sites with an exponentially decaying envelope from both sides of the maxima, consistent with bulk-boundary correspondence. Knowledge of the positions of maxima is vital for experimental realizations. Shifted MZMs have previously been observed inside vortices \cite{volovik1999fermion, read2000paired, ivanov2001non}, at defect sites \cite{fu2021majorana}, and in hybrid junctions \cite{pandey2023majorana}, where localization is tied to explicit inhomogeneities. In our work, the behavior arises from interference among multiple contributing solutions of the recursion relation, indicating that nontrivial spatial structures can emerge even in completely uniform systems. 

In experiments, the energy is not strictly zero, except in special cases, as the system size is finite. The edge-mode properties thus depend sensitively on the system size. We therefore identify parameter-dependent length scales that control finite-size behavior and connect the ideal semi-infinite results to realistic systems. Within this framework, we also clarify the distinction between zero-energy modes, edge modes, and Majorana zero modes in finite chains. We analyze how zero modes evolve to MZMs with an increase in system size, from finite chains~\cite{mahyaeh2018zero, miao2018majorana} to near-infinite limits, and show that energy splittings between edge modes govern their localization and overlap. In finite chains, edge modes typically occur at small but finite energies due to hybridization between boundary states and therefore do not fully exhibit ideal Majorana properties; this behavior arises purely from finite-size effects ~\cite{dourado2025twositekitaevsweetspots}. Such modes, often referred to as massive edge modes ~\cite{Viyuela_2016, Alecce_2017}, have been extensively studied in the literature.
 
The paper is structured as follows: Section~\ref{modelsys} outlines the model Hamiltonian, its Nambu-spinor form, and the phase diagram. Section~\ref{majoranabases} presents the Hamiltonian in the Majorana basis and derives the recursion relation. Section~\ref{roots} explains how the roots of the characteristic equation characterize phase transitions. Section~\ref{edgetypes} discusses the different types of edge modes and the conditions under which they form. Section~\ref{sizemode} examines finite size effects. Section~\ref{discussion} presents the discussion and conclusions. Additional details are provided in the appendices.

\section{Model System}
\label{modelsys}
The Kitaev chain ~\cite{Kitaev2001} with nearest and next-nearest-neighbor interactions is described by the Hamiltonian ~\cite{Alecce_2017}
\begin{equation}
	\label{kitaevGeneral}
		H = -\sum_{j=1}^{L} \mu'(c_j^{\dagger}c_j - \frac{1}{2})+ \sum_{l=1}^{2} \sum_{j=1}^{L-l} (-t_l c_j^{\dagger} c_{j+l} + \Delta_l c_j c_{j+l} + h.c)
\end{equation}
Here, $\mu'$, $t_l$, and $\Delta_l$ denote the chemical potential, hopping amplitudes, and pairing amplitudes, respectively. If $\mu' = -2\mu$ and $t_i = \Delta_i = \lambda_i$ for $i = 1, 2$, Eq.~\ref{kitaevGeneral} reduces to a special subclass of the generalized Kitaev model, which we consider throughout this work ~\cite{sticlet2013, mahyaeh2018zero, Niu_2012}:

\begin{equation}
	\label{Eq.Hamiltonian_Fermion_Basis}
		H=-\mu \sum_{j=1}^N (1-2c_j^{\dagger}c_j) -\lambda_1\sum_{j=1}^{N-1}(c_{j}^{\dagger}c_{j+1} + c_{j}^{\dagger}c_{j+1}^{\dagger} + h.c) - \lambda_2\sum_{j=2}^{N-1}(c_{j-1}^{\dagger}c_{j+1} + c_{j+1}c_{j-1} + h.c)
\end{equation}	
Here, $\mu$, $\lambda_1$, and $\lambda_2$ represent the on-site chemical potential, nearest-neighbor, and next-nearest-neighbor couplings, respectively. This Hamiltonian is also equivalent to the fermionic representation of the quantum Ising chain with an additional three-spin interaction ~\cite{Niu_2012}. Throughout this paper, unless stated otherwise, the term Kitaev model refers specifically to the Hamiltonian given in Eq.~\ref{Eq.Hamiltonian_Fermion_Basis}.

\vspace{0.2cm}

\noindent \textbf{Bulk spectrum and phase boundaries:}

To determine the bulk properties, we impose periodic boundary conditions and diagonalize the Hamiltonian using a Bogoliubov transformation. This yields an effective two-level Hamiltonian of the form

\begin{equation}
    H=\vec{\chi}\cdot\vec{\sigma}=\chi_z(k)\sigma_z - \chi_y(k)\sigma_y
\end{equation}
where $\sigma_i$ are Pauli matrices, and the coefficients are given by 
\[
\chi_z(k) \equiv -2\lambda_1\cos{k} - 2\lambda_2\cos{2k} + 2\mu, \quad
\chi_y(k) \equiv 2\lambda_1\sin{k} + 2\lambda_2\sin{2k}.
\]

The corresponding excitation spectrum is~\cite{Niu_2012}
\begin{equation}\label{spectra}
    \epsilon_k = \pm\sqrt{\chi_y^2(k) + \chi_z^2(k)}
\end{equation}

The condition $\epsilon_k = 0$ determines the bulk gap-closing points, which signal a topological phase transition. This leads to the critical lines $\lambda_2 = \mu + \lambda_1$, $\lambda_2 = \mu - \lambda_1$, and $\lambda_2 = -\mu$ (see Appendix~\ref{critical lines}). The locations of these gap-closing points depend on the system parameters like $\mu$, $\lambda_1$, and $\lambda_2$. 

\vspace{0.2cm}

\noindent \textbf{Open boundary conditions and BdG formulation:}

For open boundary conditions, translation invariance is broken, and edge modes may arise due to a nonequivalent local environment at the boundaries instead of the bulk. To analyze this case, we introduce the Nambu spinor~\cite{Nambu1960}
\[
\ket\psi = \begin{pmatrix} 
    c \\[0.1em] 
    c^\dagger
    \end{pmatrix},
\]
which allows the Hamiltonian to be written in Bogoliubov-de-Gennes (BdG) form as 
\[
H = \bra\psi H'\ket\psi, \quad 
H' \equiv \begin{pmatrix}
    \hat{h} & \hat{\Delta} \\[0.5em] 
    \hat{\Delta}^\dagger & -\hat{h}
    \end{pmatrix}.
\]

The matrices $\hat{h}$ and $\hat{\Delta}$ encode the hopping and pairing structure of the system and are given by
\begin{align}
\begin{split}
\hat{h}_{ij} = &-\frac{\lambda_1}{2} (\delta_{j,i+1} + \delta_{j,i-1}) - \frac{\lambda_2}{2} (\delta_{j,i+2} + \delta_{j,i-2}) + \mu\delta_{ij}, \\
\hat{\Delta}_{ij} =  &-\frac{\lambda_1}{2} (\delta_{j,i+1} - \delta_{j,i-1}) - \frac{\lambda_2}{2} (\delta_{j,i+2} - \delta_{j,i-2}).
\end{split}
\end{align} 

The coefficients and certain signs differ from those used in Refs.~\cite{rahul2021majorana, Niu_2012} due to a choice of convention. However, these differences do not affect the spectrum, the particle-hole symmetry, or the existence of MZMs. The explicit factors of $1/2$ are retained to facilitate comparison between different basis conventions.

\vspace{0.2cm}

\noindent \textbf{BdG eigenvalue problem and zero modes:}

The BdG eigenvalue equation reads
    
\begin{equation}
    \begin{pmatrix}
    \hat{h} & \hat{\Delta} \\[0.5em] 
    \hat{\Delta}^\dagger & -\hat{h}
    \end{pmatrix}
    \begin{pmatrix}
    \vec{u}_n \\[0.5em]
    \vec{v}_n
    \end{pmatrix}
    = E_n
    \begin{pmatrix}
    \vec{u}_n \\[0.5em] 
    \vec{v}_n
    \end{pmatrix}
\end{equation}

\begin{figure}[htbp]
\centering
\begin{subfigure}{0.8\linewidth}
\includegraphics[width=\linewidth]{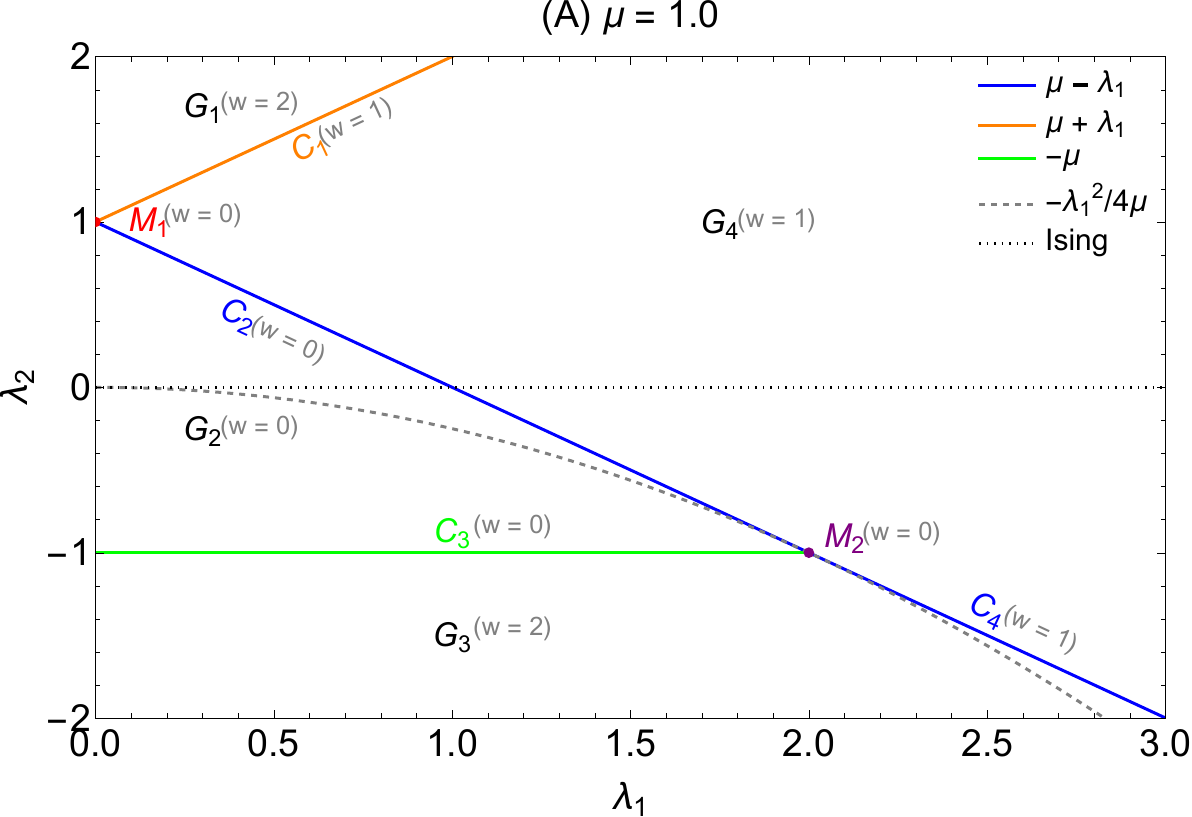}    
\end{subfigure}

\vspace{0.3cm}

\begin{subfigure}{0.8\linewidth}
\includegraphics[width = \linewidth]{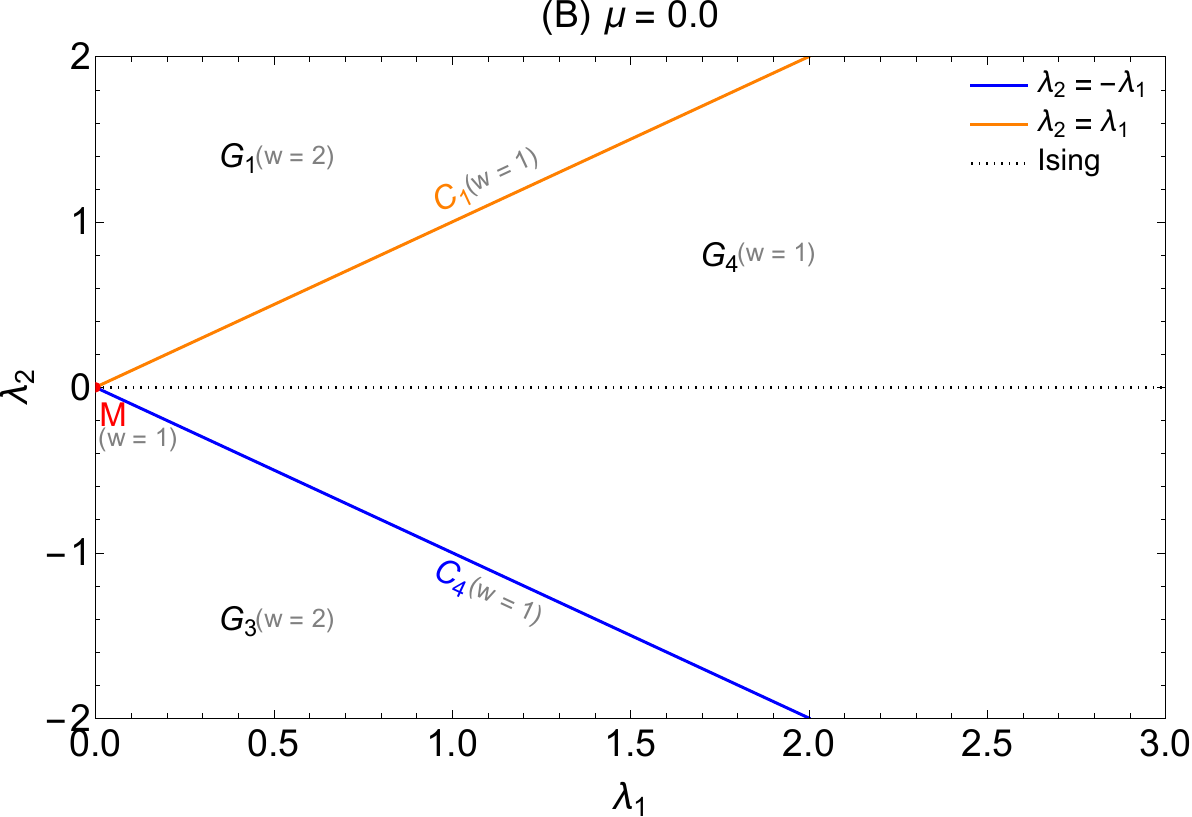}
\end{subfigure}    
\caption{\justifying{Phase diagram in the $(\lambda_1, \lambda_2)$ plane showing the number of MZMs ($\text{w}$) in different parameter regimes for fixed chemical potential $\mu$. Regions labeled $G_i$ denote distinct gapped phases characterized by different numbers of normalizable MZMs, while $C_i$ indicate the corresponding phase boundaries where the energy gap closes. The dashed curve marks the boundary separating regions with real and complex characteristic roots $q$ (see Sec.~\ref{majoranabases}).
Panel (A) corresponds to $\mu = 1$, while panel (B) corresponds to $\mu = 0$. As $\mu$ is reduced, the region in which the characteristic roots $q$ are complex shrinks and disappears entirely at $\mu = 0$. Consequently, the $\text{w} = 0$ region ($G_2$), the critical line $C_3$, and the dashed curve present in panel (A) are absent from the phase diagram at $\mu = 0$.
}}
    \label{phase}
\end{figure}

Because the Hamiltonian in Eq.~\ref{modelsys} is real and possesses an effective time-reversal symmetry in addition to particle-hole symmetry, it belongs to symmetry class BDI, for which the topological invariant is an integer winding number. In topological phases, this eigenvalue problem admits normalizable zero-energy solutions, with the number of MZMs equal to the winding number~\cite{Niu_2012}. At a topological phase transition, where the bulk gap closes, zero modes from the adjoining phases combine such that the number of normalizable MZMs is given by the minimum of those in the adjoining phases. The number of normalizable MZMs ($\text{w}$) in different parametric regimes is shown in Fig.~\ref{phase}, which also illustrates how the extent of the topological regions depends on the chemical potential ($\mu$). Due to particle-hole symmetry, if $(\vec{u}_n, \vec{v}_n)$ is an eigenstate with energy $E_n$, then $(\vec{v}_n^*, \vec{u}_n^*)$ is an eigenstate with energy $-E_n$. At zero energy, both correspond to the same eigenvalue, and one can choose linear combinations that satisfy the self-conjugacy condition $\vec{u}_n = \vec{v}_n^*$, corresponding to Majorana modes. In the present case, where the Hamiltonian is real, the eigenvectors can be chosen real, reducing this condition to $\vec{u}_n = \pm \vec{v}_n$. The probability density of an eigenmode with energy $E_n$ at site $j$ is given by $P_j=|u_{n,j}|^2+|v_{n,j}|^2$. 

All quantities used throughout this work are dimensionless, with energies expressed in units of the characteristic coupling parameters of the Hamiltonian.

\section{Physics from Majorana bases}
\label{majoranabases}

To analyze the spatial structure of zero modes beyond what is captured by topological invariants, it is convenient to rewrite the Hamiltonian in the Majorana basis. If we convert the fermionic basis operators, $c_i$ and $c_i^\dagger$, into Majorana basis operators $a_i$ and $b_i$, the Hamiltonian (Eq.~\ref {Eq.Hamiltonian_Fermion_Basis}) can be rewritten as (see Appendix~\ref{MajoranaBasis}). 

\begin{equation}\label{FermionH}
H = -2i\left[-\mu\sum_{i=1}^N b_ia_i + \lambda_1\sum_{i=1}^{N-1} b_ia_{i+1} +  \lambda_2\sum_{i=2}^{N-1}b_{i-1}a_{i+1}\right]
\end{equation}

The normalized operators are defined as $ a_i = \dfrac{c^\dagger_i + c_i}{\sqrt{2}}$ and $b_i = -\dfrac{i(c^\dagger_i - c_i)}{\sqrt{2}}$
such that $\{a_i, a_j\} = \delta_{i,j} = \{b_i, b_j\}$ and $\{a_i, b_j\} = 0$. This normalization ensures that the energy scale remains identical in both the fermionic and Majorana representations, and no rescaling of constant terms is required, unlike in some earlier conventions~\cite{Niu_2012}. 

Choosing the basis $(a_1, b_1, a_2, b_2, \cdot\cdot\cdot, a_n, b_n)$, the eigenvector corresponding to energy $E$ can be obtained from the coupled recursion relations (Appendix~\ref{RecursionSolution})
\begin{subequations}\label{Majoranaeigenvector}
    \begin{align}
    -\mu A_j + \lambda_1A_{j+1} + \lambda_2A_{j+2} = & iE A_{j}^{'} \\
    \mu A_{j}^{'} - \lambda_1A_{j-1}^{'} - \lambda_2A_{j-2}^{'} = &  iE A_{j}
    \end{align}
\end{subequations}
with boundary conditions arising from the absence of sites beyond the chain, namely $A_{0}^{'} = 0 = A_{-1}^{'}$ and $A_{N+1} = 0 = A_{N+2}$. The corresponding eigenvector is of the form $(A_1, A_1^{'}, A_2, A_2^{'}, \cdot\cdot\cdot, A_N, A_N^{'})^T$. To obtain any eigenmode, these coupled equations must be solved simultaneously along with the boundary conditions. If we attempt to find a mode at the exact $E=0$ in a finite chain, we obtain only the trivial solution. This can be seen explicitly: for $j=1$, the second equation reduces to $\mu A_1^{'} = i E A_1$, while for $j=N$, the first equation yields $-\mu A_N = i E A_N^{'}$. Setting $E=0$, together with the boundary conditions, enforces $A_N=0=A_1^{'}$, irrespective of the values of $A_1$ and $A_N^{'}$. Since $A_1^{'}$ and $E$ are exactly zero, the second equation forces all $A'$ amplitudes to vanish. Similarly, since $A_N = 0$, the first equation forces all $A$ amplitudes to vanish, leading to the trivial solution. This outcome follows from the structure of the Hamiltonian, the boundary conditions, and the requirement of exact zero energy in a finite chain, and does not indicate any breakdown of the recursion approach. 

The above discussion shows that an exact MZM cannot exist in a finite chain unless the mode is perfectly localized. In that case, all $A'$ amplitudes vanish, while $A$ is non-zero only over a finite number of sites near the left end, so that the boundary conditions at the opposite end are automatically satisfied. In more general situations, the energy is not exactly zero but very small. In such cases, the amplitudes $A'$ become very small due to the left boundary conditions, while $A$ decays sufficiently fast so that it does not extend significantly toward the opposite boundary. As a result, the boundary conditions are effectively satisfied, yielding a left-localized mode that is not perfectly localized to a single or two sites. This occurs when the system size is much larger than the zero-mode decay length. In practice, such small but non-zero energies are more commonly observed than exact zero modes in finite systems. To capture these idealized zero-mode solutions, we consider the limit of a semi-infinite chain, where the right boundary condition is no longer relevant. In this limit, one can consistently set all $A'$ amplitudes to zero, and the solution is determined entirely by $A$, which exhibits an overall decaying behavior. The zero-energy recursion relation then reduces to
\begin{equation}\label{recursion}
    -\mu A_j + \lambda_1A_{j+1} + \lambda_2A_{j+2} = 0
\end{equation}
The corresponding eigenvector takes the form $(A_1, 0, A_2, 0, \cdot\cdot\cdot,A_N,0)^T$, with no additional boundary conditions on $A$. To solve this recursion relation, we assume the ansatz $A_i=q^i$, which yields the general solution
\begin{equation}
    A_j = C_1 q_+^j + C_2 q_-^j
\end{equation}
where $q_\pm$ are the roots of the characteristic equation $\lambda_2 q^2 + \lambda_1 q - \mu = 0$, given by 
\[
q_{\pm} = \dfrac{-\lambda_1 \pm \sqrt{\lambda_1^2 + 4 \mu \lambda_2}}{2 \lambda_2}.
\]

The characteristic roots of the recursion relation provide a unified analytical description of the MZMs: their existence, localization length, and spatial behavior are determined directly by the roots' magnitudes and phases. In particular, the magnitude controls the localization length, while the relative phase determines whether the decay is monotonic or oscillatory and the positions of the probability maxima. Thus, analysis of the characteristic equation establishes a direct connection between the Hamiltonian parameters and the spatial profiles of the MZMs across the phase diagram. Once the roots are known, key features of the MZMs, including their localization, decay pattern, and possible shifts of the probability maximum, can be understood without requiring numerical diagonalization of finite systems, provided the energy is sufficiently low to use the recursion relation given by Eq.~\ref{recursion}. The probability density at site $j$ is $ P_j = |A_j|^2$.

Although the Nambu representation formally doubles the eigenvectors due to particle-hole symmetry of the Bogoliubov-de Gennes Hamiltonian, the physically relevant solutions correspond to MZMs localized at the system boundaries. The recursion relation generates normalizable solutions at one of the two boundaries of the chain. In particular, the ansatz $A_i=q^i$ describes modes localized at the left end when $|q|<1$. The number of such independent normalizable solutions determines the number $w$ of MZMs at that parametric point. The corresponding modes localized at the opposite end can be obtained by applying the inverse spatial dependence $A_i\propto q^{-i}$, which effectively inverts the direction of the recursion and involves only non-zero $A'$ amplitudes. Thus, the present construction captures boundary-localized modes without requiring explicit treatment of the opposite edge. The localization length or penetration depth of MZMs is set by the magnitude of the characteristic roots. In practice, the recursion relation in Eq.~\ref{recursion} is valid even when the energy is sufficiently close to zero for a finite chain whose effective cut-off length is set by this penetration depth. The normalizable solutions, therefore, correspond to the roots satisfying $|q_\pm|<1$. 

Based on this criterion (Appendix~\ref{w}), the number of MZMs in different regions is shown in Fig.~\ref{phase}. If both roots are normalizable, the general solution can be written in terms of the amplitudes at the first two sites, $A_1$ and $A_2$ (Appendix~\ref{w2generalsolution}):

\begin{align}\label{Generalw2}
    A_j = \dfrac{1}{q_+ - q_-}[A_1(q_-^{j-1} q_+ - q_-q_+^{j-1}) + A_2(q_+^{j-1} - q_-^{j-1})]
\end{align}

This solution is valid only when $q_+\neq q_-$. While both $A_1$ and $A_2$ appear as free parameters, the overall normalization fixes one degree of freedom, so that only their relative values determine a given zero-mode profile.

Two independent choices of the relative coefficients yield two independent MZMs, provided the corresponding vectors are orthogonal. If they are not, an orthogonal mode can be constructed using:
\begin{equation*}
    \vec{B} = \vec{A}_2 - \dfrac{\braket{\vec{A}_2|\vec{A}_1}}{\braket{\vec{A}_1|\vec{A}_1}}\vec{A}_1 
\end{equation*}
where $\vec{A}_1$ and $\vec{A}_2$ are obtained from two distinct choices of the relative coefficients. The vector $\vec{B}$ is orthogonal to vector $\vec{A}_1$, and being a linear combination of valid solutions, it also satisfies the recursion relation. In the $\text{w} = 1$ region, only one root satisfies $|q|<1$, and the normalized solution reduces to 
\begin{equation*}
    A_j = c q^j
\end{equation*}
where $q$ is the normalizable root, and the overall coefficient $c$ is fixed by the overall normalization of the mode. For $\text{w} = 0$, neither root is normalizable and hence no zero-energy eigenvector exists.

Although the recursion relation appears independent of the system size, it is derived under the assumption of zero energy, which itself depends on the system size. If the size is not sufficiently large, the energy deviates from zero, and the recursion relation does not yield the correct MZMs. However, once this applicability condition is satisfied, the amplitudes $A_m$ become independent of the total system size; in particular, the coefficients for the first $k$ sites remain unchanged even if the recursion relation is extended to $m>k$.

\subsection{Phase transitions determined from the roots (q) of the characteristic equation }
\label{roots}

\begin{figure}[htbp]
    \centering
    \begin{minipage}[b]{0.32\textwidth}
        \centering
        \includegraphics[scale = 0.43]{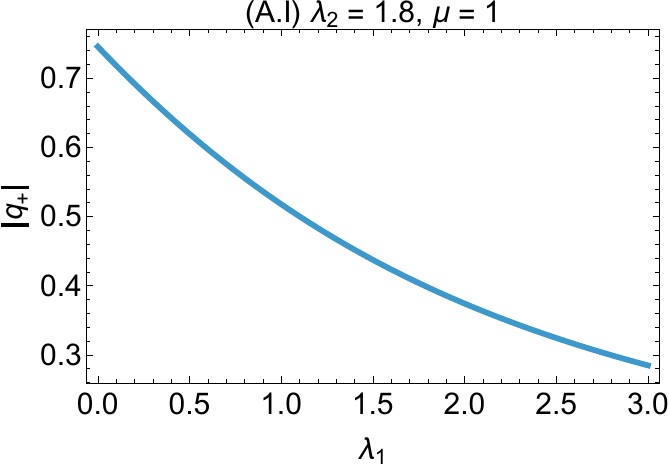}
    \end{minipage}
   % \hspace{0.1cm}
    \begin{minipage}[b]{0.32\textwidth}
        \centering
        \includegraphics[scale = 0.43]{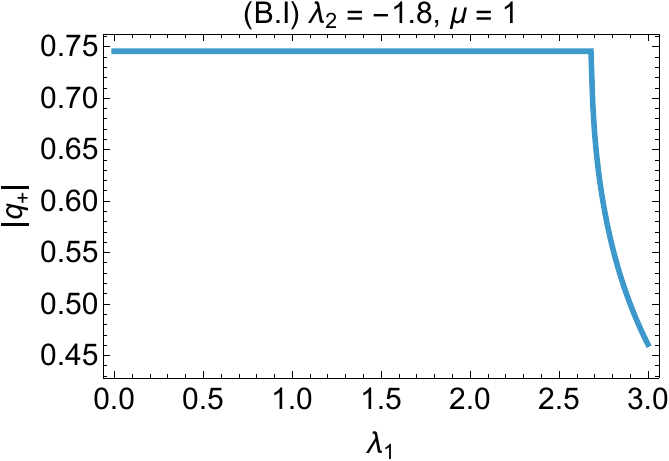}
    \end{minipage}
    %\hspace{0.1cm}
     \begin{minipage}[b]{0.32\textwidth}
        \centering
        \includegraphics[scale = 0.43]{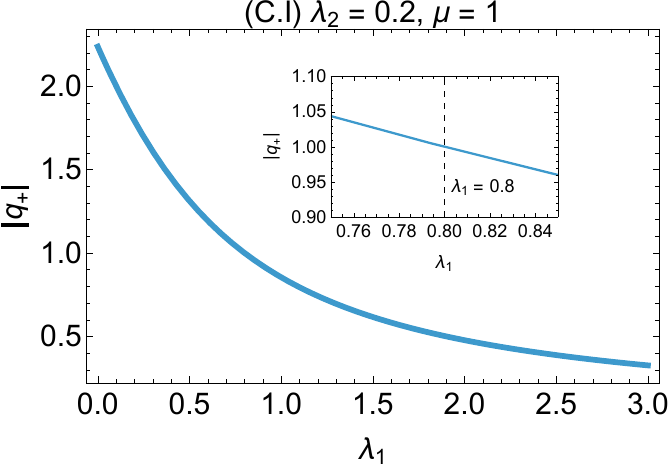}
    \end{minipage}
    
    \vspace{0.8em}  % space between rows
    
    % Second row
    \begin{minipage}[b]{0.32\textwidth}
        \centering
        \includegraphics[scale = 0.43]{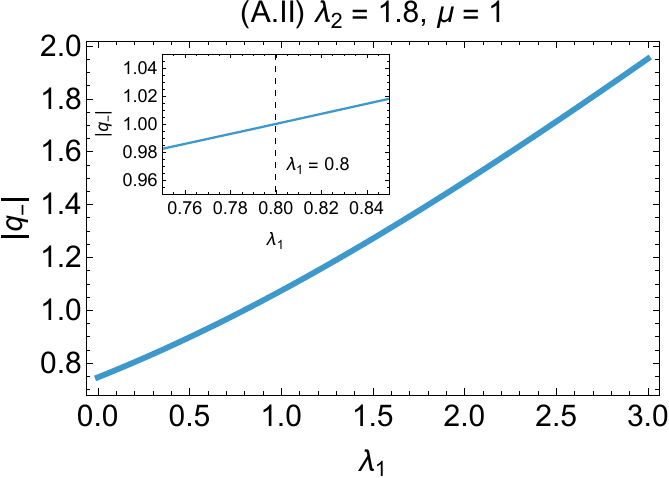}
    \end{minipage}
    %\hspace{0.1cm}
    \begin{minipage}[b]{0.32\textwidth}
        \centering
        \includegraphics[scale = 0.43]{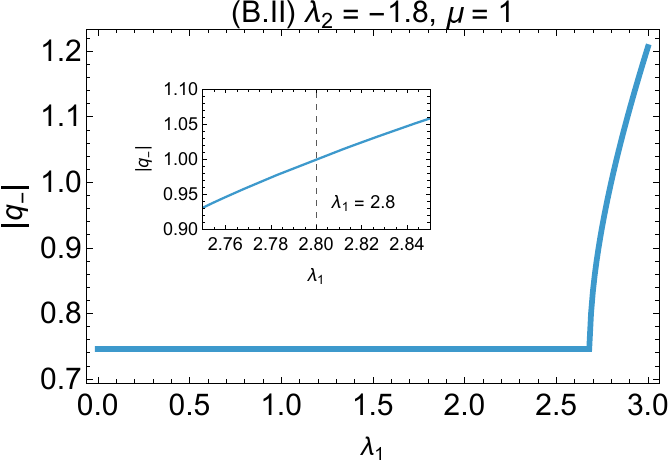}
    \end{minipage}
   % \hspace{0.1cm}
    \begin{minipage}[b]{0.32\textwidth}
        \centering
        \includegraphics[scale = 0.43]{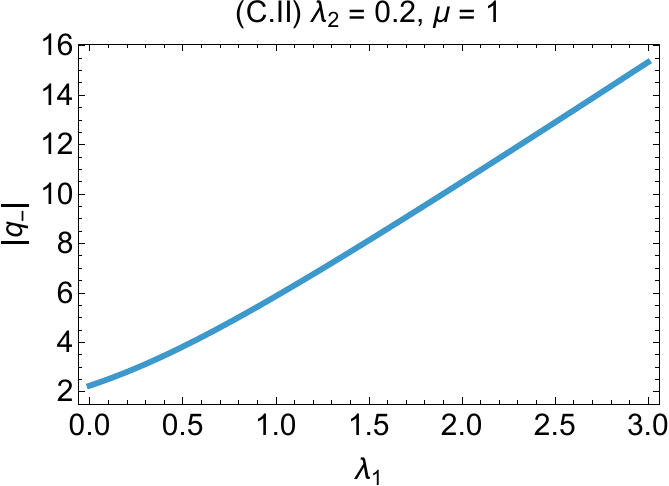}
    \end{minipage}
    \caption{\justifying{Plots of the modulus of the characteristic roots $|q_\pm|$ as a function of $\lambda_1$ for different values of $\lambda_2$ at fixed $\mu$, illustrating topological phase transitions. Panels (A.I, B.I, and C.I) show the behavior of $q_+$, while panels (A.II, B.II, and C.II) show the behavior of $q_-$. Regions with two, one, or zero normalizable roots ($|q|<1$) correspond to different topological phases with distinct numbers of Majorana zero modes.
    Panels (A.I and A.II) show the transition from region $G_1$ to region $G_4$ across the critical line $C_1$; panels (B.I and B.II) show the transition from $G_3$ to $G_4$ across $C_4$; and panels (C.I and C.II) show the transition from $G_2$ to $G_4$ across $C_2$, as indicated in Fig.~\ref{phase} (A). These transitions correspond to distinct changes in the number of roots lying inside the unit circle. In each case, the dashed line (shown in the insets) marks the corresponding critical point at which $|q|=1$, indicating a gap-closing condition and a topological phase transition. This demonstrates that the characteristic roots provide a unified analytical criterion for identifying phase transitions and tracking their proximity in parameter space.}}
    \label{q_graph}
\end{figure}

The condition for a normalizable solution is $|q|<1$. Since $\lambda_1 \geq 0$, whenever there is only one MZM, it corresponds to the root satisfying $|q_-|\geq 1$, while the normalizable solution is determined by $q_+$. All transition lines and points, therefore, correspond to $|q|=1$ for the previously normalizable mode. Within a given phase, the unnormalizable always satisfies $|q| > 1$. Unlike the winding number, which remains constant within a phase, modulus $|q|$ varies continuously across the parameter space. As a result, it not only signals phase transitions but also provides a measure of the distance from the corresponding critical lines or points. When the roots are real, $|q|$ varies smoothly across regions, as shown in Fig.~\ref{q_graph} (A.I and A.II). However, when the roots transition from complex to real, the variation becomes non-smooth, as seen in Fig.~\ref{q_graph} (B.I and B.II). For $\lambda_1^2 + 4 \mu \lambda_2 < 0$, the roots are complex and satisfy
\[
|q_\pm| = \sqrt{\frac{-\mu}{\lambda_2}},
\] 
which is independent of $\lambda_1$. As a result, $|q_\pm|$ remains constant as $\lambda_1$ varies, until the system reaches the dashed curve in Fig.~\ref{phase}, defined by $\lambda_1^2 + 4 \mu \lambda_2 = 0$. At this point, $\mu = \dfrac{-\lambda_1^2}{4\lambda_2}$. Substituting this into the above expression gives,
\[
|q_\pm| = \sqrt{\dfrac{\lambda_1^2}{4 \lambda_2^2}} = \dfrac{|\lambda_1|}{2|\lambda_2|}.
\]
On the dashed curve, the roots become equal, 
\[
q_+ = q_- = \dfrac{-\lambda_1}{2\lambda_2},
\]
which yields the same value
\[
|q_\pm| = \dfrac{|\lambda_1|}{2|\lambda_2|}.
\]

This shows that the transition from complex to real roots is continuous, although not smooth. The behavior of $|q|$ across different regions is illustrated in Fig.~\ref{q_graph}. In Fig.~\ref{q_graph} (A.I and A.II), $|q_+|$ remains less than one, while $|q_-|$ increases with increasing $\lambda_1$ for fixed $\lambda_2$ and $\mu$. For $\lambda_1 < 0.8$, the system lies in the $G_1$ region; at $\lambda_1=0.8$, it reaches the transition line $C_1$ where $|q_-| = 1$; for $\lambda_1 > 0.8$, $|q_-| > 1$, and the system enters the $G_4$ region, with $|q_-|$ continuing to increase with $\lambda_1$. A similar trend is observed in Fig.~\ref{q_graph} (B.I and B.II). In the regime where the roots are complex (below the dashed curve in Fig.~\ref{phase} (A)), both moduli $|q_+|$ and $|q_-|$ are equal and remain less than one. As $\lambda_1$ increases, this value remains constant until $\lambda_1\approx2.68$, corresponding to a point on the dashed curve. Beyond this point, the roots become real: $|q_{+}|$ continues to decrease, while $|q_-|$ increases. However, $|q_-|<1$ until $\lambda_1 = 2.8$, where the transition line $C_4$ is reached with $|q_-| = 1$. For larger values of $\lambda_1$, $|q_-| > 1$, and the system enters the $G_4$ region. This behavior reflects the transition from $G_3$ to $G_4$ through the transition line $C_4$.

In Fig.~\ref{q_graph} (C.I and C.II), $|q_-|$ remains greater than one and increases with $\lambda_1$, while $|q_+|$ decreases from greater than one to less than one, crossing unity at $\lambda_1=0.8$. This corresponds to a transition from the $G_2$ region to the $G_4$ region through the transition line $C_2$.

A similar continuous variation of $|q|$ is observed when varying $\lambda_2$ at fixed $\lambda_1$ and $\mu$. At $\lambda_2 = 0$, corresponding to the conventional Kitaev chain, the expression for $|q_\pm|$ appears ill-defined. However, $q_+$ takes an indeterminate $\frac{0}{0}$ form with a well-defined limiting value $\dfrac{\mu}{\lambda_1}$, whereas $q_-$ diverges. This indicates that only one normalizable zero mode exists. The same conclusion follows directly from the recursion relation $\lambda_1 A_{j+1} = \mu A_j$ (Eq.~\ref{recursion}), which yields the same value of $q$. 

The continuous variation of $|q|$ therefore provides a quantitative measure of the distance from the phase transition. Large values of $|q|$ correspond to points farther from the critical lines, while values approaching unity indicate proximity to a transition. In regions where $|q|$ remains constant, the system is effectively equidistant from the transition line $C_3$ until the dashed curve is reached; beyond this point, $|q|$ begins reflecting the change in the nature of the roots.

\subsection{Different types of zero modes}
\label{edgetypes}
The root analysis above determines the localization properties of the zero modes from the magnitudes and phases of the characteristic roots. We now examine representative parameter regimes of the phase diagram and illustrate how these root structures manifest in the spatial profiles of the MZMs.

Edge modes are commonly identified as states whose probability density is maximal at the boundary and decays toward the bulk. This criterion is straightforward when only a single MZM is present. However, the situation becomes more subtle when multiple MZMs exist. In such cases, the spatial structure of the modes can differ significantly, and the site of maximum probability need not coincide with the boundary. The examples presented below illustrate how the characteristic roots of the recursion relation control this behavior. 

A simple example, discussed in detail later, arises when two perfectly localized MZMs are present. Since the two MZMs cannot occupy the same lattice site, one MZM may be localized at the first site, while the other appears at the neighboring sites. In this situation, the first MZM is naturally identified as an edge MZM, while the second MZM lies further inside the chain. More generally, there are cases where the maximum probability of a boundary-origin MZMs is located well inside the lattice. Although the maximum may occur away from the boundary, the envelope decays exponentially on either side of the peak. Despite this shift, the MZM remains boundary-origin as its structure is still determined by the boundary conditions rather than by bulk localization.

Depending on the system parameters and system size, several types of zero modes can arise: (a) overlapping left and right modes. (b) perfectly localized left or right modes, and (c) perfectly localized modes that also satisfy Majorana conditions. Within these categories, the spatial behavior may vary, with modes exhibiting monotonic decay, oscillatory decay, or perfect localization. In finite systems, there are also cases where the energy is not exactly zero, yet the mode still has its largest probability near the edge and decays toward the bulk. 

In the following, we focus on the semi-infinite limit, in which the intrinsic structure of the MZMs can be analyzed without finite-size effects, such as overlap between left and right modes. Organizing the discussion in terms of representative parameter regimes also allows direct identification of parameter choices that produce specific types of mode profiles. We now analyze different cases for parameters within individual phases in this limit.

\subsubsection{\texorpdfstring{$\mu\neq0$, $\lambda_1=0$, $\lambda_2=0$}{}}
This parameter set corresponds to a trivial limit of the conventional Kitaev model in which only the chemical potential ($\mu$) is present. All lattice sites are effectively decoupled as the hopping and pairing terms vanish. The recursion relation reduces to $\mu A_j = 0$ (Eq.~\ref{recursion}), which requires all $A_j$ to vanish for a zero-energy solution. Hence, no MZM exists in this case. This point lies at $(0, 0)$ in the region $G_2$ of Fig.~\ref{phase} (A).

\subsubsection{\texorpdfstring{ $\mu=0$, $\lambda_1\neq0$, $\lambda_2=0$}{}}
 
\begin{figure}[htbp]
    \centering
        \includegraphics[scale = 0.43]{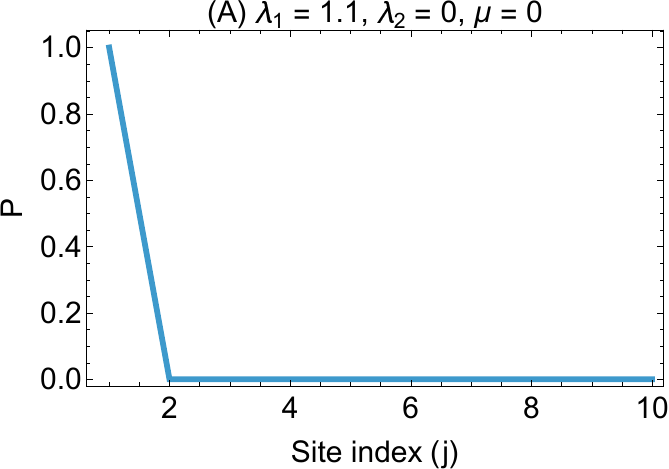}
        \hspace{0.01cm}
        \includegraphics[scale = 0.43]{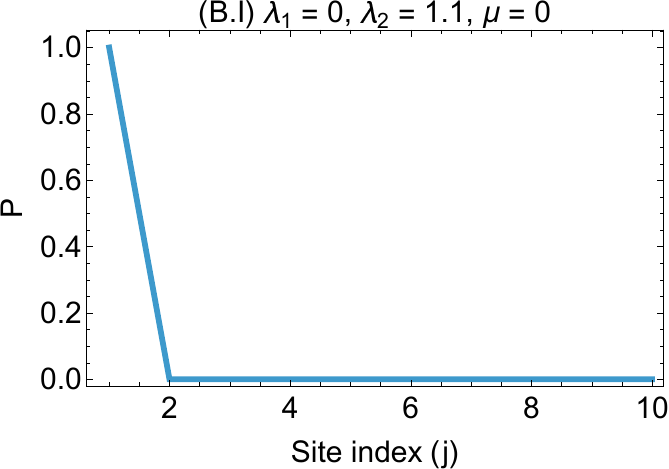}
        \hspace{0.01cm}
        \includegraphics[scale = 0.43]{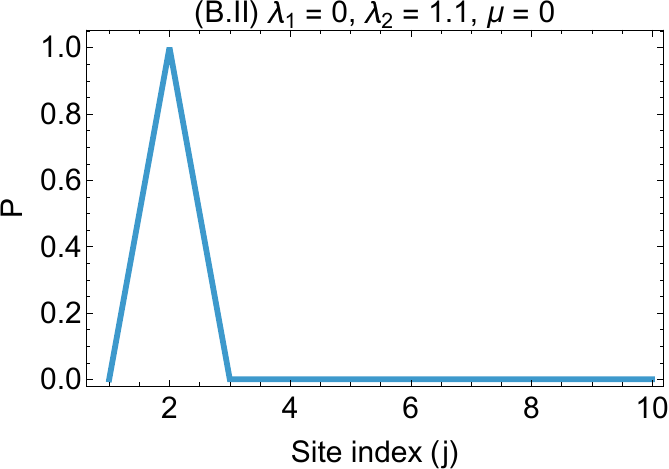}
   
    \caption{\justifying{Probability distribution (P) as a function of lattice site $j$. (A) Single MZM for $\lambda_2 = 0$ and $\mu=0$ with $\lambda_1 \neq 0$, corresponding to the nearest-neighbor Kitaev model, where the MZM is perfectly localized at the first site. (B.I and B.II) Two MZMs for $\mu = 0$ and $\lambda_1 = 0$ with $\lambda_2 \neq 0$. In this case, both MZMs are perfectly localized, with probability concentrated at the first and second sites, respectively. These perfectly localized profiles arise in special parameter regimes in which the characteristic roots yield strictly localized solutions without exponential tails. All MZMs shown have exact zero energy.}}
     \label{mu0L10}
\end{figure}

This is a special case of the Kitaev model. In the phase diagram for $\mu=0$ (Fig.~\ref{phase} (B)), it corresponds to a dotted line lying entirely within the $\text{w} = 1$ phase, indicating the presence of a single MZM for all $\lambda_1$. In this limit, the extreme left Majorana operator $a_1$ and the extreme right Majorana operator $b_N$ are absent from the Hamiltonian, and hence correspond to exact MZMs as they do not appear in any coupling terms. As a result, these MZMs are perfectly localized at the first and last sites, respectively, and their existence is independent of the system size. The same conclusion follows from the recursion relation $\lambda_1 A_{j+1} = 0$ (Eq.~\ref{recursion}), which forces all amplitudes $A_j$ to vanish except $A_1$, which is fixed by normalization. This corresponds to a perfectly localized MZM at the first site. This case is illustrated in Fig.~\ref{mu0L10} (A).

\subsubsection{\texorpdfstring{$\mu=0$, $\lambda_1=0$, $\lambda_2\neq0$}{}}

This case corresponds to the y-axis in Fig.~\ref{phase} (B). There are two MZMs for all $\lambda_2 \neq 0$; at $\lambda_2=0$, the Hamiltonian vanishes, and the system becomes ill-defined. The recursion relation reduces to $\lambda_2 A_{j+2}=0$ (Eq.~\ref{recursion}), which implies that all amplitudes vanish beyond the first two sites. The solution is therefore determined by the coefficients $A_1$ and $A_2$, which correspond to two perfectly localized MZMs at the first and second sites, respectively, as shown in Fig.~\ref{mu0L10} (B.I and B.II). More general solutions correspond to linear combinations of these modes, characterized by different relative values of $A_1$ and $A_2$, and can have support on both sites while remaining localized to the first two lattice sites.

\subsubsection{\texorpdfstring{$\mu=0$, $\lambda_1\neq 0$, $\lambda_2\neq0$}{}}

\begin{figure}[htbp]
    \centering
        \includegraphics[scale = 0.55]{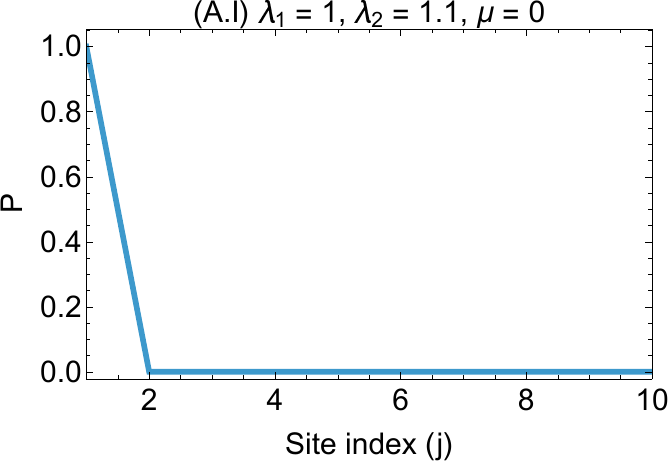}
        \hspace{1.2cm}
        \includegraphics[scale = 0.55]{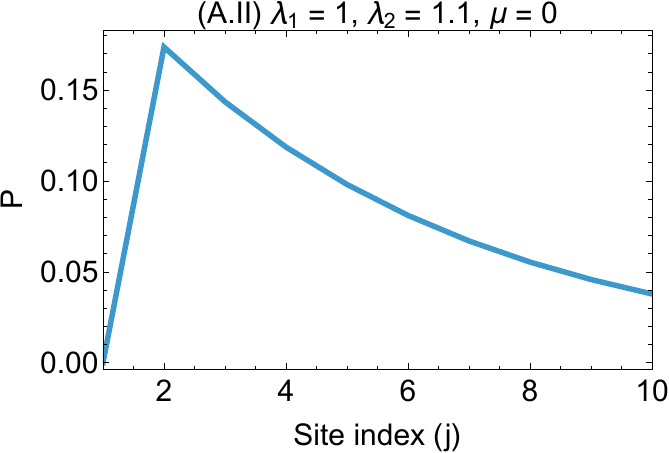}
    \caption{\justifying{Probability distribution (P) as a function of lattice site $j$ for the extended Kitaev chain at $\mu=0$. Panel (A.I) shows a perfectly localized MZM. Panel (A.II) shows the second MZM, which is orthogonal to the first and spatially extended across the chain, with amplitudes that decay away from the boundary.}}
    \label{mu0}
\end{figure}

Fig.~\ref{phase} (B) shows that this parameter regime contains regions with neither $\text{w}=0$ nor complex roots. In this case, $q_+ = 0$, independent of $\lambda_1$ and $\lambda_2$. The recursion relation reduces to (Eq.~\ref{recursion})
\[
\lambda_2 A_{j+2} + \lambda_1A_{j+1}=0 
\]
with a general solution 
\[
A_{j} = A_2\bigg(\dfrac{-\lambda_1}{\lambda_2}\bigg)^{j-2}
\]
which is independent of $A_1$ for $j\geq 2$. The solution space is therefore determined by the coefficients $A_1$ and $A_2$, corresponding to the amplitudes at the first two sites. 

The normalizability condition requires $\left|\frac{\lambda_1}{\lambda_2}\right| < 1$, which defines the $\text{w}=2$ regions. In this case, two linearly independent MZMs exist. One MZM is obtained by setting $A_2 = 0$, for which the amplitudes vanish for all $j\geq 2$, yielding a perfectly localized MZM at the first site. After normalization, this corresponds to $A_1 = 1$. The second MZM corresponds to $A_2 \neq 0$, for which the amplitudes extend into the bulk, producing a decaying profile. More general solutions correspond to linear combinations of these two MZMs, characterized by different relative values of $(A_1, A_2)$. Fig.~\ref{mu0} (A.I and A.II) shows these two cases explicitly: a perfectly localized MZM at the first site and an orthogonal MZM with finite amplitude extending into the bulk. 

In the $\text{w}=1$ region, the condition $|q_-|\geq 1$ reduces the solution to $A_1=1$ and $A_j=0$ for $j\geq2$, corresponding to a perfectly localized MZM at the first site. Thus, perfectly localized MZMs arise in both the $\text{w}=1$ and $\text{w}=2$ regions as a consequence of the structure of the recursion relation. This behavior differs from that of the Kitaev model without $\lambda_2$, in which the localization properties are determined solely by the parameters.

This regime has also been explored experimentally~\cite{tenHaaf2025}. Although that work considers only a three-site system ($N=3$), the choice $\mu=0$ with equal pairing and hopping amplitudes maps their Hamiltonian onto the parameter regime of the present model. The considered parameters lie within our $\text{w}=2$ region (Fig.~\ref{phase} (B)), where one MZM remains perfectly localized and independent of system size, while the second MZM exhibits a decaying tail toward the bulk. In practice, observing this second MZM requires a system size larger than its decay length, consistent with the behavior shown in Fig.~\ref{mu0} (A.II).

\subsubsection{\texorpdfstring{$\mu\neq0$, $\lambda_1\neq0$, $\lambda_2=0$}{}}

This limit corresponds to the standard nearest-neighbor Kitaev chain where a single MZM exists for $\lambda_1 > \mu$. The same condition follows directly from the recursion relation (Eq.~\ref{recursion}),
\[
\lambda_1 A_{j+1} = \mu A_j
\]
whose solution is 
\[
A_j = \bigg(\frac{\mu}{\lambda_1}\bigg)^{j-1}A_1.
\]
The solution is normalizable when $|\mu / \lambda_1| < 1$, with $A_1$ fixed by normalization. This parameter regime corresponds to the black dotted line in Fig.~\ref{phase}(A).

\subsubsection{\texorpdfstring{$\mu\neq0$, $\lambda_1 = 0$, $\lambda_2\neq0$}{}}

This case corresponds to the y-axis of Fig.~\ref{phase} (A), excluding the origin. Along this line, the number of MZMs changes from two to zero and back to two as $\lambda_2$ is increased. The recursion relation reduces to (Eq.~\ref{recursion})
\[
\lambda_2A_{j+2} = \mu A_j.
\]
The corresponding characteristic equation has roots 
\[
q_\pm = \pm\sqrt{\frac{\mu}{\lambda_2}} \quad (\lambda_2 > 0),\qquad
q_\pm = \mp i\sqrt{\frac{\mu}{|\lambda_2|}} \quad (\lambda_2 < 0).
\]
In both cases, the roots have equal magnitudes but differ only in sign or phase. Using $q_+ = -q_-$ in Eq.~\ref{Generalw2}, the general solution can be written in the unified form 
\[
A_j = \frac{1}{2q_+}\big[A_1 q_+^j((-1)^{j-1} + 1) + A_2q_+^{j-1}(1 - (-1)^{j-1}))\big],
\]
where the value of $q_+$ depends on the sign of $\lambda_2$. 

\begin{figure}[htbp]
    \centering
        \includegraphics[scale = 0.55]{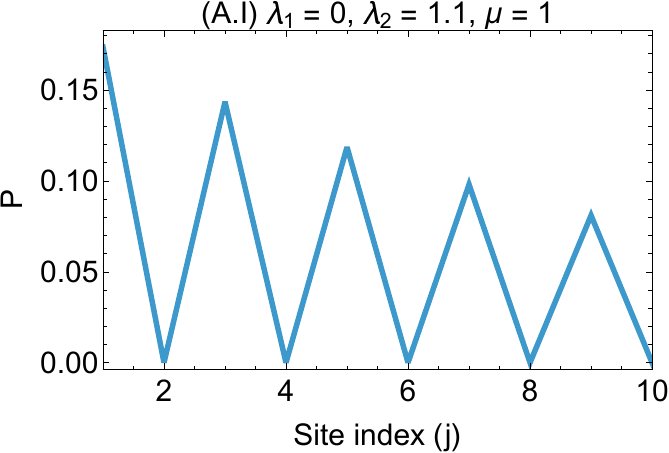}
        \hspace{1.2cm}
        \includegraphics[scale = 0.55]{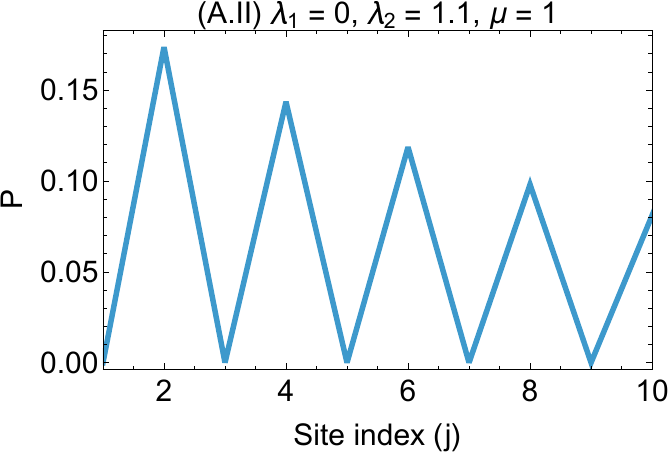}
    \caption{\justifying{Probability distribution (P) as a function of lattice site $j$ for the extended Kitaev chain at $\lambda_1=0$. Both panels show spatially extended MZMs. The two MZMs correspond to different choices of the relative amplitudes of the independent solutions (e.g., $A_1=1$ and $A_2=0$ in (A.I) and $A_1=0$ and $A_2=1$ in (A.II)), which enforces vanishing amplitudes on alternating lattice sites. As a result, the MZMs have support exclusively on either the odd or even sublattices.}}
    \label{lam10}
\end{figure}

This expression shows that the amplitudes decouple across sublattices: on odd sites, only the $A_1$ contribution survives, whereas on even sites, only the $A_2$ contribution remains. For special cases in which one of the coefficients vanishes ($A_1=0$ or $A_2=0$), the MZMs have non-zero weight exclusively on either the odd or the even sublattice, with an overall decaying envelope, as shown in Fig.~\ref{lam10}. When both coefficients are non-zero, the amplitudes exhibit oscillatory decay. The probability density attains its maxima at either the first or the second site, depending on the relative values of $(A_1, A_2)$.

\subsubsection{\texorpdfstring{$\mu\neq0$, $\lambda_1\neq 0$, $\lambda_2\neq0$}{}}

\begin{figure}[htbp]
        \centering
        \includegraphics[scale=0.55]{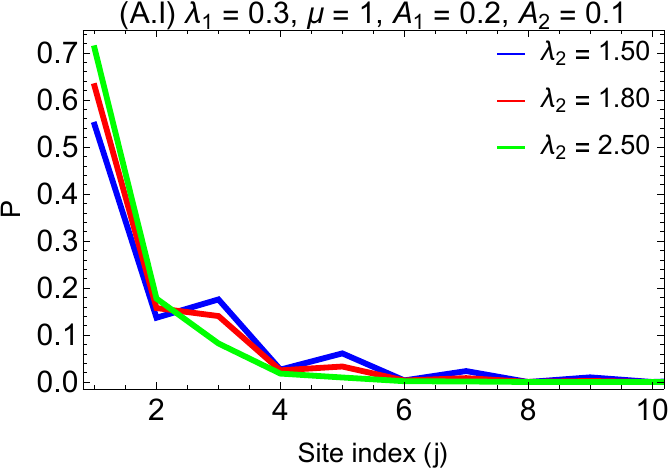}
        \hspace{1.2cm}
        \includegraphics[scale=0.55]{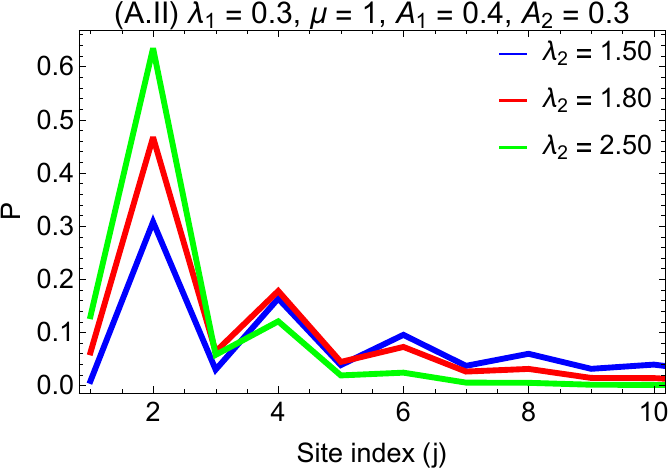}
     \caption{\justifying{Probability distributions illustrating the evolution between oscillatory and monotonic decay. (A.I) MZM profile showing a transition from oscillatory decay to monotonic decay as the parameters are varied. (A.II) The corresponding orthogonal MZM remains oscillatory decay. For certain parameter choices (e.g., $\lambda_2 = 2.50$), the orthogonal MZM exhibits oscillatory decay even when the primary MZM exhibits monotonic decay. These results show that, within this parameter regime, the MZMs can either both exhibit oscillatory decay or display distinct behaviors, with one oscillatory decay and the other monotonic decay, depending on the system parameters and the relative amplitudes of $(A_1, A_2)$.}}
     \label{recursionG1}
 \end{figure}

This case is shown in Fig.~\ref{phase} (A). While all regions except the y-axis and the line $\lambda_2 = 0$ satisfy the normalizability condition, the form of the characteristic roots $q$ differs significantly across parameter regimes. As a result, the spatial structure of MZMs varies even though the number of MZMs and the winding number remain constant within a given phase. These behaviors can be classified into five distinct cases:

\textit{Case 1} \textbf{Both roots are real, unequal, and normalizable:} 
The general solution (Eq.~\ref{Generalw2}) can be recast as (Appendix~\ref{w2realq})
\begin{align}\label{realqunequal}
    A_j = q_-A_{j-1} + (A_2 - q_-A_1)q_+^{j-2}
\end{align}
A closed-form expression is 
\begin{equation}
    A_j = q_-^{j-2}A_2 + \sum_{k=1}^{j-2} f(k)q_-^{j-k-2}
\end{equation}
where $f(k)=(A_2 - A_1q_-)q_+^{k}$. 

This case occurs in the region $G_1$ and part of the region $G_3$ where $\lambda_1^2 + 4\mu\lambda_2 > 0$, $\lambda_2 < \mu-\lambda_1$, and $\lambda_2 < -\mu$.  

\textbf{Region $G_1$:} Here $\lambda_1 > 0$, $\lambda_2 > 0$ for $\mu>0$, so that $ q_- < 0$ and $q_+ > 0$. For positive initial conditions $A_1$ and $A_2$, the term $(A_2 - q_{-}A_1)q_+^{j-2}$ remains positive. The amplitude $A_j$ is therefore determined by the competition between this positive term and the contribution from $q_{-}$, which can lead to alternating signs depending on their relative magnitudes. As a result, the spatial profile can exhibit either oscillatory or monotonic decay Fig.~\ref{recursionG1}. 

As $\lambda_2$ increases, the magnitude of $q_-$ increases while $q_+$ decreases. This reduces the imbalance between the competing terms and suppresses sign alternation, thereby diminishing oscillatory behavior. For smaller $\lambda_2$, the contrast between these contributions is stronger, resulting in more pronounced oscillations. This behavior is further influenced by the relative values of $(A_1, A_2)$, which determine the coefficient $(A_{2} - q_{-}A_{1})$ and hence the relative weight of the contributing terms. Consequently, orthogonal MZMs, corresponding to different relative weightings, can exhibit distinct spatial behaviors. 

In the special case $A_2 = q_{-} A_1$, the shifting term vanishes and the recursion reduces to $A_j = q_{-}A_{j-1}$. Although the sign of $A_j$ alternates, the probability decays monotonically as $|A_j|^2 = |q_-|^2|A_{j-1}|^2$. 

In this region, the probability maximum is confined to the first or the second site, irrespective of whether the MZM exhibits oscillatory decay or monotonic decay. This follows from the recursion relation: 
\[
A_3 = \frac{-\lambda_1}{\lambda_2}A_2 + \frac{\mu}{\lambda_2}A_1,
\]
where both $\frac{\lambda_1}{\lambda_2}$ and $\frac{\mu}{\lambda_2}$ are less than unity. As a result, $A_3$ cannot exceed both $A_1$ and $A_2$, ensuring that the maximum remains within the first two sites. For larger $j$, the contribution from the shifting term (Eq.~\ref{realqunequal}) decreases further due to the factor $q_+$, while the contribution from $q_-$ remains unchanged at each step. This further suppresses amplitudes away from the boundary, maintaining the localization of the maximum near the first or second site. 

\textbf{Region $G_3$ with $\lambda_1^2 + 4\mu\lambda_2 > 0$ and $\lambda_2 < -\mu$:}
In this regime, the signs of $q_-$ and $q_+$ are reversed compared to region $G_1$, such that $q_- > 0$ and $q_+ < 0$. The reducing factor $q_-$ therefore leads to a monotonic decrease in magnitude, while the shifting term $(A_2 - q_- A_1) q_+^{j-2}$ alternates in sign and decreases in magnitude due to the factor $q_+$. As in region $G_1$, the interplay between these two contributions determines whether the MZM exhibits oscillatory decay or monotonic decay. Despite the difference in sign structure, the overall qualitative behavior of the MZMs remains similar to that in region $G_1$
 
\vspace{0.3 cm}

\begin{figure}[htbp]
	\centering
		\includegraphics[scale = 0.55]{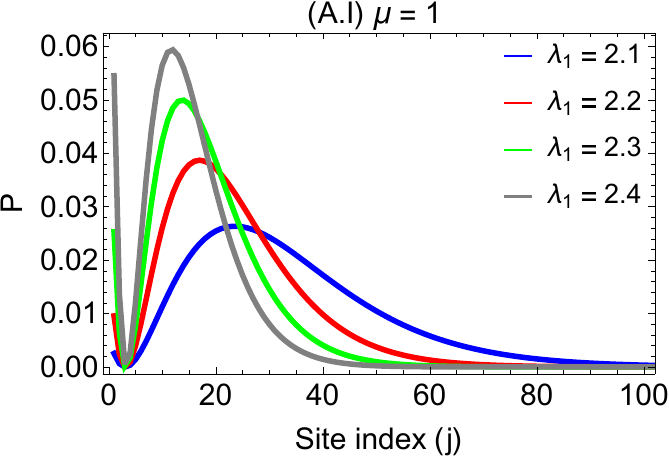}
	\hspace{1.2cm}
		\includegraphics[scale = 0.55]{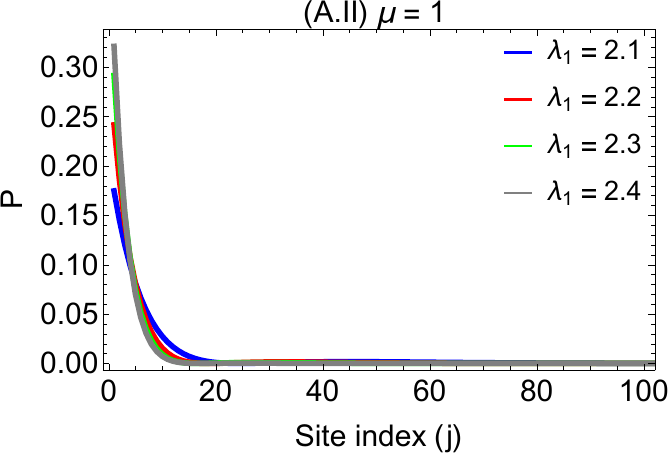}
	\caption{ \justifying{Probability distribution for the case of degenerate characteristic roots $q_+ = q_-$, corresponding to the dashed curve $\lambda_2 = -\lambda_1^2/4\mu$ in Fig~\ref{phase}(A). (A.I) MZM profile illustrating the shift of the site of maximum probability as $\lambda_1$ is varied. For $\lambda_1 = 2.1$, the probability maximum occurs deep inside the lattice interior, while the MZM remains boundary-origin and exhibits an exponentially decaying envelope on both sides of the maximum. The profile additionally shows oscillatory decay despite the roots being real. (A.II) The corresponding orthogonal MZM, which exhibits purely monotonic decay. These profiles demonstrate that the position of maximum probability is governed by the characteristic roots and can shift deep into the lattice interior without destroying the boundary-origin nature of MZMs.}}
	\label{recursiondashed}
\end{figure}

\textit{Case 2} \textbf{Both roots are real, equal, and normalizable:} 
For equal roots, Eq.~\ref{Generalw2} is not applicable; instead, the solution takes the form 
\[
A_j = (c_1+c_2 j)r^j
\]
which can be written in terms of $A_1$ and $A_2$ as (Appendix~\ref{w2equalq})
\[
A_j = [(2-j) q A_1 + (j-1)A_2]q^{j-2}
\]
Although $q$ is real, the interplay between the linear prefactor $(2-j) q A_1 + (j-1)A_2$ and the exponential term $q^{j-2}$ can shift the probability maximum into the bulk. The dependence of the peak position on $\lambda_1$ is shown in Fig.~\ref{recursiondashed}. This case occurs along the dashed line $\lambda_1^2 + 4\mu\lambda_2 = 0$ in Fig.~\ref{phase} (A).

\vspace{0.3cm}

\textit{Case 3} \textbf{Both roots complex, unequal, and normalizable:} 
This case occurs only in region $G_3$, where $\lambda_2 < 0$ and $\lambda_1^2 + 4\mu\lambda_2 < 0$ (for $\mu > 0$). The solution can be written as (Appendix~\ref{w2complexq})
\begin{equation}
    A_j = \dfrac{R^{j-2}}{\sin\theta}\,y\,\sin{\big[(j-1)\theta - \phi\big]}
\end{equation}
where $y = \sqrt{A_1^2 R^2 + A_2^2 + 2A_1 A_2 R\cos{\theta}}$, $\cos{\phi} = \frac{A_1 R \cos{\theta}}{y}$, $R=\sqrt{\frac{-\mu}{\lambda_2}}$, and $\cos\theta=\frac{\lambda_1}{\sqrt{-4\mu\lambda_2}}$.

\begin{figure}[htbp]
     \centering
        \includegraphics[scale = 0.55]{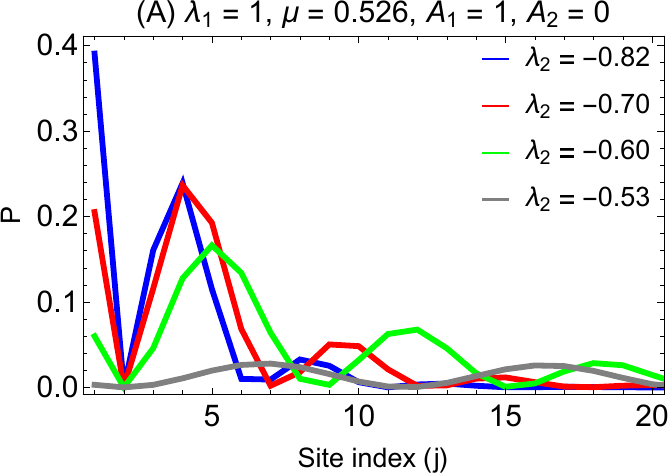}
       \hspace{1.2cm}
        \includegraphics[scale = 0.55]{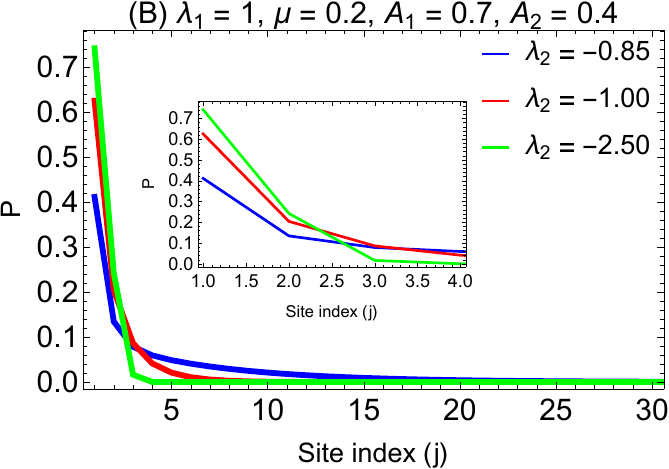}
     \caption{ \justifying{Variation of the site of maximum probability under parameter changes. (A) Shift of the probability peak for MZMs exhibiting oscillatory decay. (B) MZM profiles displaying purely monotonic decay with different decay rates. These results show that, even within the same parameter region where the characteristic roots $q$ are complex, MZMs do not necessarily exhibit oscillatory decay. Instead, the specific spatial profile depends sensitively on the system parameters and the relative amplitudes of $(A_1, A_2)$.}}
     \label{recursionG3}
 \end{figure}

In this regime, the probability maximum can lie deep within the bulk, with its position tunable by system parameters (Fig.~\ref{recursionG3}). The MZMs may exhibit either oscillatory decay or monotonic decay and are qualitatively distinct from bulk states. To illustrate these features clearly, $\mu$ is chosen to differ from unity so that $\lambda_2$ does not attain excessively large values. Varying $\mu$ modifies the extent of the topological regions, thereby allowing a clearer visualization of the different behaviors, as shown in Fig.~\ref{phase}(B).

\vspace{0.3cm}

\textit{Case 4} \textbf{Only one root normalizable:} 
The solution reduces to $A_j = c q_+^j$, where $c$ is fixed by normalization. This MZM is purely monotonically decaying. This case corresponds to \textbf{region $G_4$}. Complex roots do not arise here since $|q_{+}| = |q_{-}|$ for complex $q$, which would make either both or neither normalizable.

\vspace{0.3cm}

\textit{Case 5} \textbf{No root normalizable:} 
No MZM exists in this case. This occurs throughout the region \textbf{region $G_2$} of Fig.~\ref{phase} (A), despite the system being gapped. Both real and complex roots can arise within this region, depending on the parameters. The region between the lines $\lambda_2 = -\mu$ and $\lambda_2 = 0$ corresponds to complex roots, while the region above the line $\lambda_2 = 0$ corresponds to real roots. In all these cases, however, neither root satisfies the normalizability condition, and hence no MZM is supported.

\section{Finite-size effects and comparison with the semi-infinite limit}
\label{sizemode}
Although the recursion relation provides a transparent characterization of MZMs in the semi-infinite limit, experimentally realized systems are inherently finite. Only exact zero-energy Majorana modes are strictly size-independent without additional constraints. It is therefore essential to compare the analytical predictions obtained in the semi-infinite limit analysis with finite-size results~\cite{Alecce_2017, leumer2020exact, mahyaeh2018zero}. This comparison allows us to identify which spatial features of the MZMs remain observable in realistic systems. 

\vspace{0.2cm}

\noindent \textbf{Finite-size behavior and experimental relevance}

Experimental situations can be broadly classified into two categories: (1) parameters can be tuned such that well-approximated zero modes appear within accessible system sizes, and (2) tuning is not possible, in which case the finite-size features of the observed modes must be carefully analyzed. 

From the coupled finite-size eigenvalue equations, generic eigenmodes are linear combinations of both Majorana operators $a_i$ and $b_i$. Consequently, the modes are overlapping, with contributions from both ends of the chain. The threshold below which a mode can be regarded as effectively zero energy, therefore, depends not only on the topological phase but also on the acceptable degree of mode overlap and deviation from ideal Majorana character, i.e., on how much non-ideal behavior one is willing to tolerate.

\begin{figure}[htbp]
    \centering
    \hspace{-0.4cm}
    \begin{minipage}[b]{0.32\textwidth}
        \centering
        \includegraphics[scale = 0.43]{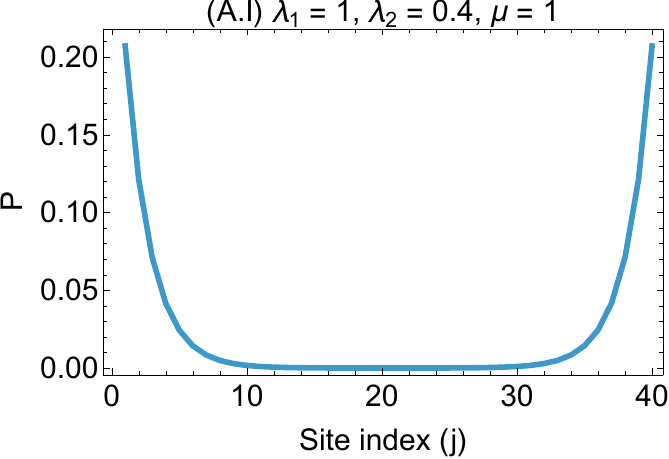}
        %\caption*{\hspace{1cm} (a)}
    \end{minipage}
    %\hspace{0.5cm}
    \begin{minipage}[b]{0.32\textwidth}
        \centering
        \includegraphics[scale = 0.43]{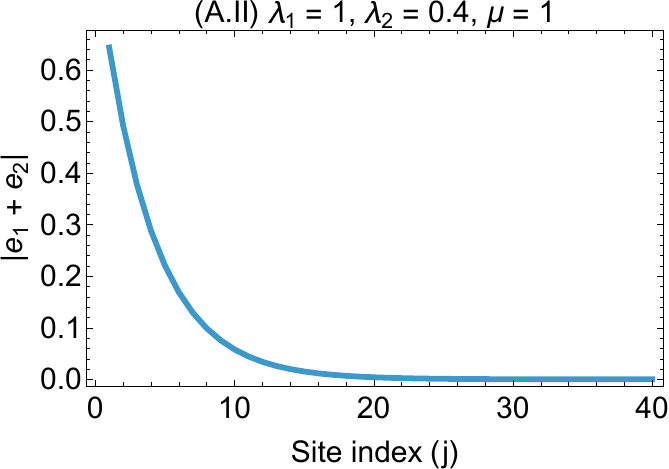}
        %\caption*{\hspace{1cm}(b)}
    \end{minipage}
    %\hspace{0.5cm}
    \begin{minipage}[b]{0.32\textwidth}
        \centering
        \includegraphics[scale = 0.43]{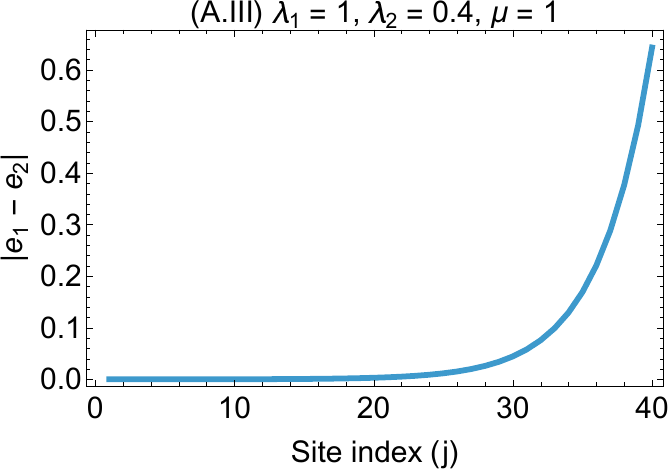}
        %\caption*{\hspace{1cm}(c)}
    \end{minipage}
    \vspace{0.2cm}
    \hspace{-0.4cm}
    \begin{minipage}[b]{0.32\textwidth}
        \centering
        \includegraphics[scale = 0.43]{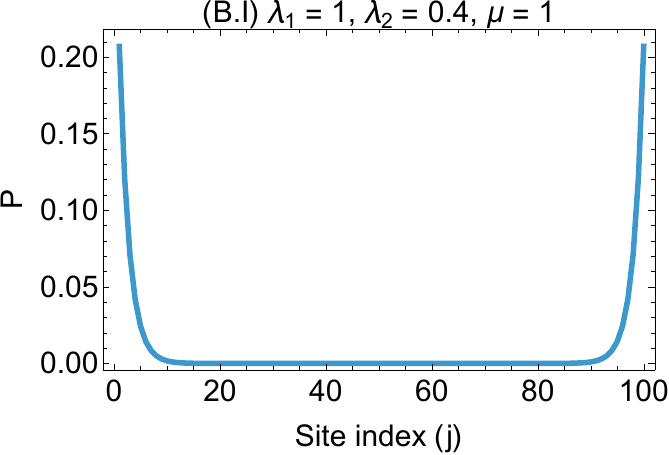}
        %\caption*{\hspace{1cm} (d)}
    \end{minipage}
    %\hspace{0.5cm}
    \begin{minipage}[b]{0.32\textwidth}
        \centering
        \includegraphics[scale = 0.43]{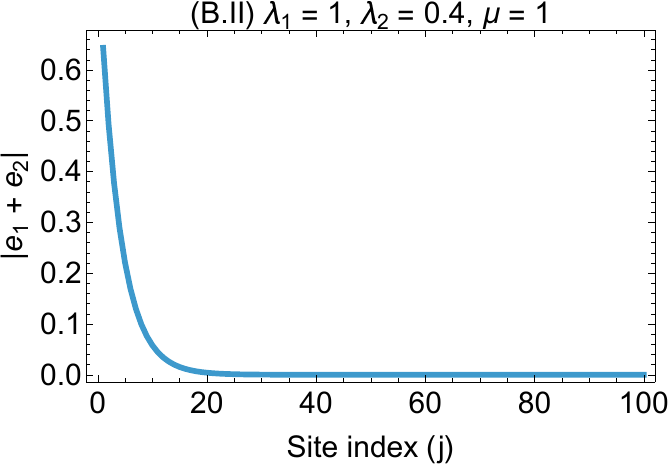}
        %\caption*{\hspace{1cm}(e)}
    \end{minipage}
   % \hspace{0.5cm}
    \begin{minipage}[b]{0.32\textwidth}
        \centering
        \includegraphics[scale = 0.43]{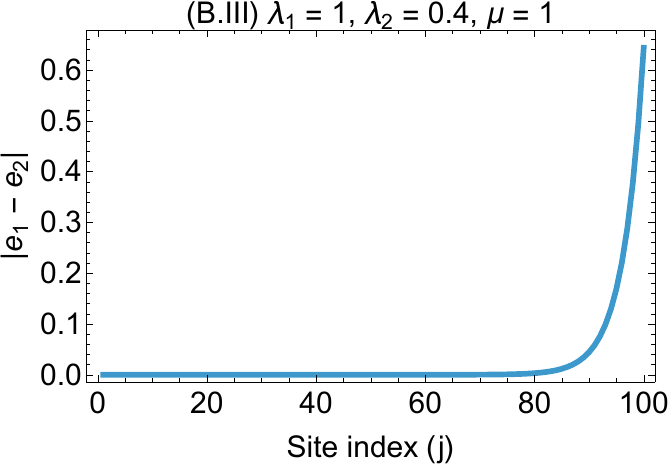}
        %\caption*{\hspace{1cm}(f)}
    \end{minipage}
    \vspace{0.2cm}
    \hspace{-0.4cm}
    \begin{minipage}[b]{0.32\textwidth}
        \centering
        \includegraphics[scale = 0.43]{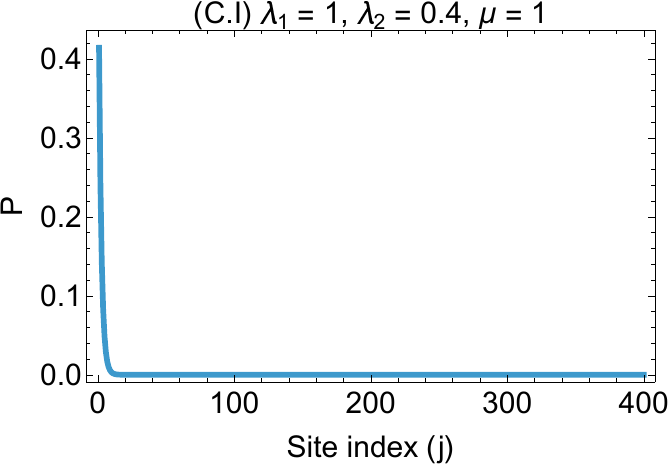}
       % \caption*{\hspace{1cm} (g)}
    \end{minipage}
   % \hspace{0.5cm}
    \begin{minipage}[b]{0.32\textwidth}
        \centering
        \includegraphics[scale = 0.43]{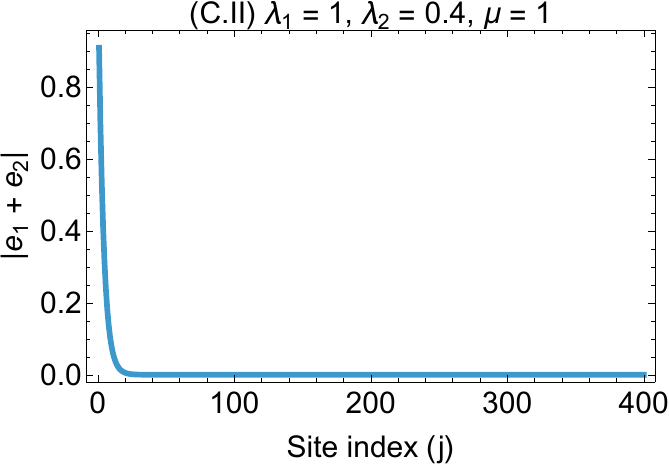}
        %\caption*{\hspace{1.5cm}(h)}
    \end{minipage}
    %\hspace{0.5cm}
    \begin{minipage}[b]{0.32\textwidth}
        \centering
        \includegraphics[scale = 0.43]{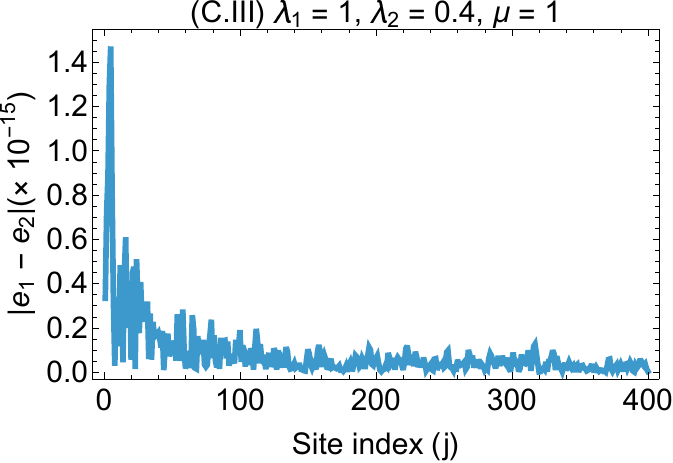}
       % \caption*{\hspace{1cm}(i)}
    \end{minipage}
    \caption{\justifying{Evolution of edge modes with system size in the $\text{w} = 1$ phase. (A.I, B.I and C.I) show the probability distributions of the lowest-energy eigenmodes for chain lengths $n=40$, $n=100$, and $n=400$, respectively. The corresponding sum and difference of the eigenvector components are shown in (A.II, B.II, and C.II) and (A.III, B.III, and C.III), respectively. Here, the eigenvector components $e_1$ and $e_2$ correspond to indices $1$ to $n$ and $n + 1$ to $2n$, respectively, representing the fermionic operators $c$ and $c^\dagger$. The Majorana operators are obtained via the basis transformation $a_i=e_1+e_2$ and $b_i=e_1-e_2$. The corresponding eigen energies for $n = 40$, $n = 100$, and $n = 400$ are $-0.0000152638$, $-1.66941\times 10^{-12}$, and $5.55112\times10^{-16}$, respectively, representing the eigenvalues closest to zero for the chosen parameters. As the system size increases, the energy approaches zero, and the initially overlapping modes progressively collapse into a single left- or right-localized edge mode. This crossover is marked by the suppression of the $a_i$ components across the lattice and a monotonically decaying profile of the $b_i$ components from one end of the chain.}}
    \label{ExIsingW1}
\end{figure}

\begin{figure}[htbp]
    \centering
    \hspace{-0.4cm}
    \begin{minipage}[b]{0.32\textwidth}
        \centering
        \includegraphics[scale = 0.43]{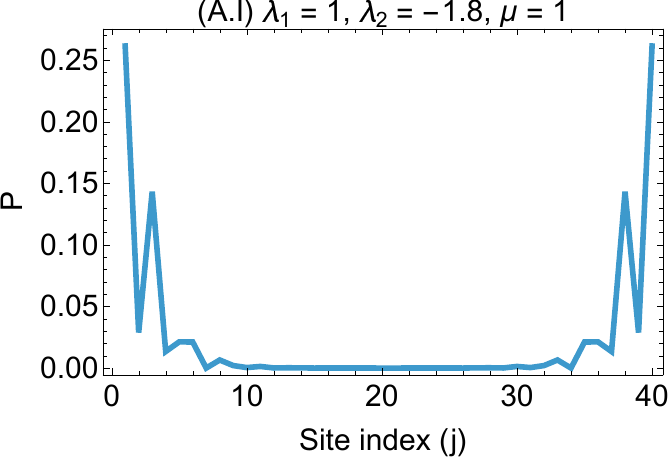}
       % \caption*{\hspace{1cm} (a)}
    \end{minipage}
    %\hspace{0.5cm}
    \begin{minipage}[b]{0.32\textwidth}
        \centering
        \includegraphics[scale = 0.43]{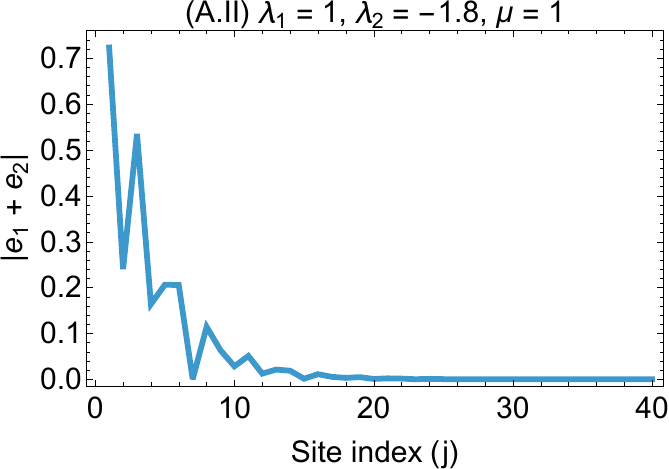}
        %\caption*{\hspace{1cm}(b)}
    \end{minipage}
    %\hspace{0.5cm}
    \begin{minipage}[b]{0.32\textwidth}
        \centering
        \includegraphics[scale = 0.43]{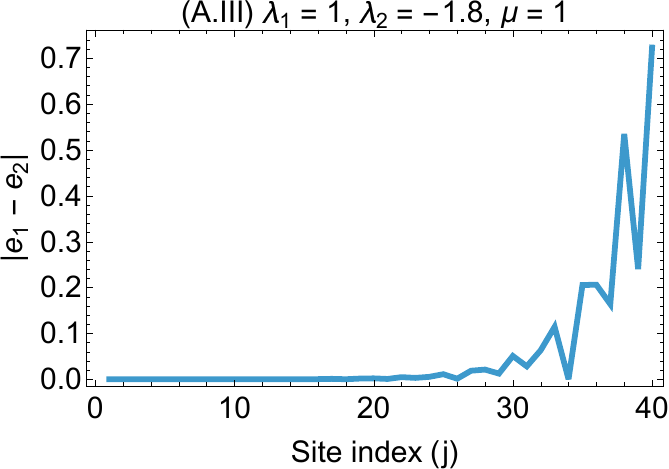}
        %\caption*{\hspace{1cm}(c)}
    \end{minipage}
    \vspace{0.2cm}
    \hspace{-0.4cm}
    \begin{minipage}[b]{0.32\textwidth}
        \centering
        \includegraphics[scale = 0.43]{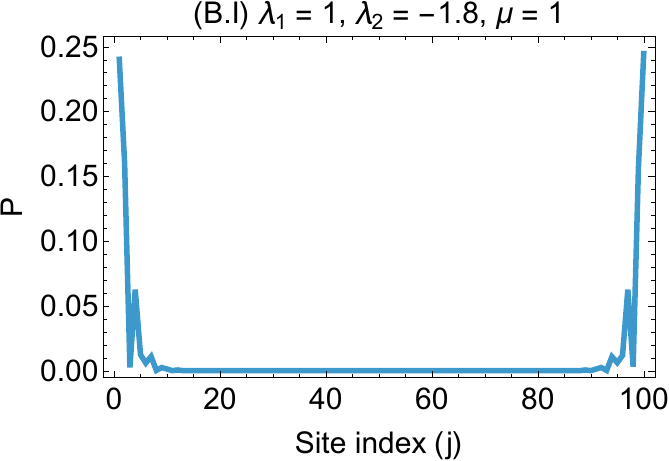}
        %\caption*{\hspace{1cm} (d)}
    \end{minipage}
    %\hspace{0.5cm}
    \begin{minipage}[b]{0.32\textwidth}
        \centering
        \includegraphics[scale = 0.43]{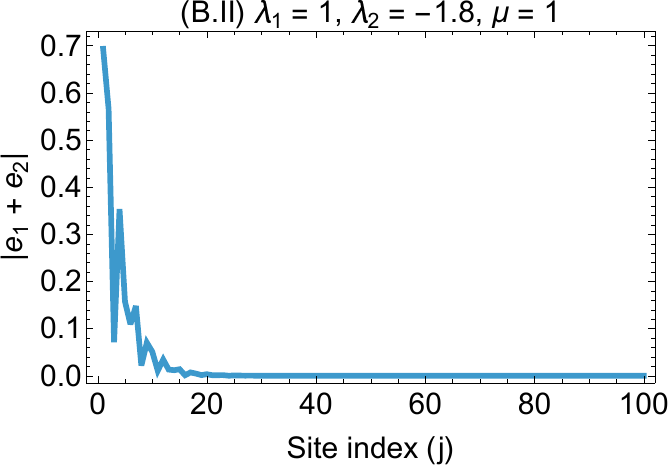}
        %\caption*{\hspace{1cm}(e)}
    \end{minipage}
    %\hspace{0.5cm}
    \begin{minipage}[b]{0.32\textwidth}
        \centering
        \includegraphics[scale = 0.43]{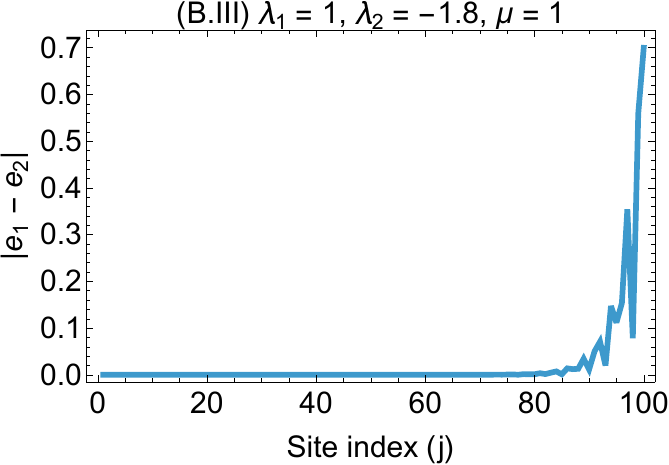}
        %\caption*{\hspace{1cm}(f)}
    \end{minipage}
    \vspace{0.2cm}
    \hspace{-0.4cm}
    \begin{minipage}[b]{0.32\textwidth}
        \centering
        \includegraphics[scale = 0.43]{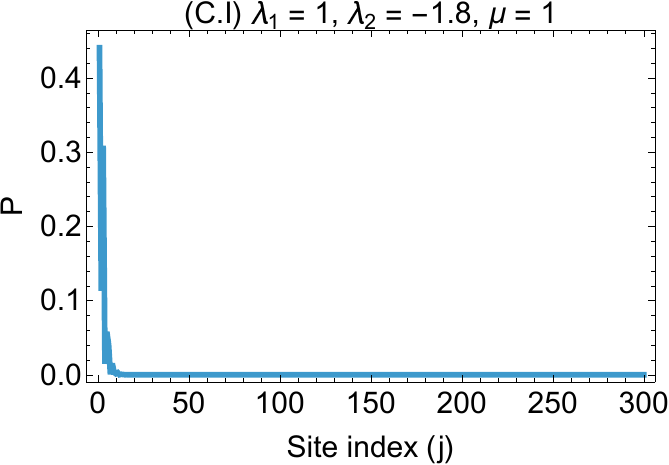}
        %\caption*{\hspace{1cm} (g)}
    \end{minipage}
    %\hspace{0.5cm}
    \begin{minipage}[b]{0.32\textwidth}
        \centering
        \includegraphics[scale = 0.43]{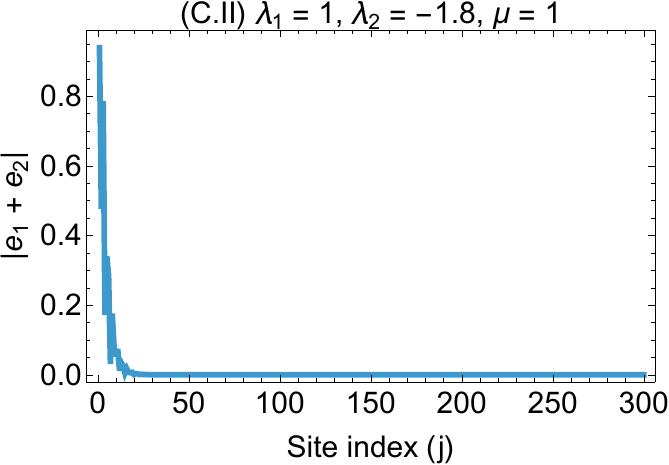}
        %\caption*{\hspace{1cm}(h)}
    \end{minipage}
    %\hspace{0.5cm}
    \begin{minipage}[b]{0.32\textwidth}
        \centering
        \includegraphics[scale = 0.43]{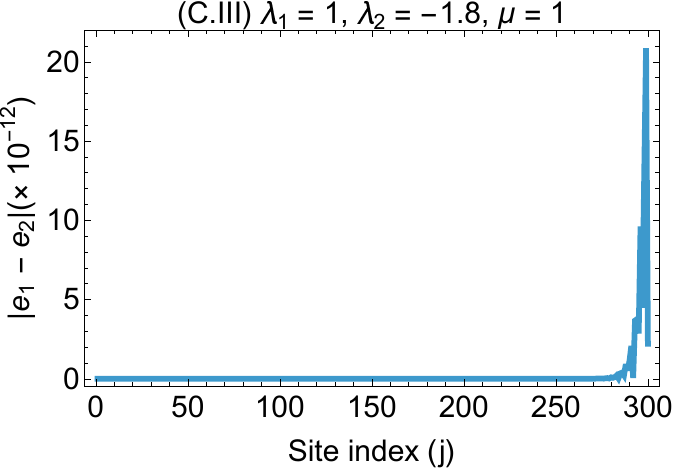}
        %\caption*{\hspace{1cm}(i)}
    \end{minipage}
    \caption{\justifying{Finite-size evolution of edge modes in the bottom $\text{w} = 2$ region. Panels (A.I, B.I, and C.I) show the probability distributions of the lowest-energy eigenmodes for chain lengths $n = 40$, $n = 100$, and $n = 300$, respectively. The corresponding sum and difference of the eigenvector components are shown by (A.II, B.II, and C.II) and (A.III, B.III, and C.III), respectively. As in Fig.~\ref {ExIsingW1}, these correspond to the Majorana components $a_i$ and $b_i$ obtained from the fermionic eigenvectors. The corresponding eigenvalues for $n = 40$, $n = 100$ and $n = 300$ are $8.65056 \times 10^{-6}$, $1.60893 \times 10^{-13}$ and $-6.84002\times10^{-16}$ respectively. A trend similar to that observed in Fig.~\ref {ExIsingW1} is evident: as the system size increases, the energy approaches zero and the initially overlapping modes progressively separate into well-separated left- and right-localized edge modes. This crossover is characterized by the suppression of one Majorana component ($a_i$ or $b_i$) across the lattice.}}
    \label{ExIsingW2}
\end{figure}

This behavior is illustrated in Fig.~\ref{ExIsingW1} and Fig.~\ref{ExIsingW2}. As shown in Fig.~\ref{ExIsingW1} (A.I, A.II, and A.III), when the energy is of order $\sim 10^{-5}$, the mode is strongly overlapping; the left-end probabilities are predominantly associated with Majorana $a$ operators, while the right-end probabilities arise from Majorana $b$ operators. 

As the system size increases, the energy decreases to $\sim 10^{-12}$. Although the mode remains overlapping, the number of sites with non-zero probability decreases, indicating a reduction in the penetration depth (Fig.~\ref{ExIsingW1} (B.I, B.II, and B.III)). For sufficiently large system sizes, when the mode becomes strongly left-localized with negligible contribution from the $b$ Majorana component, the energy gets further reduced to $\sim 10^{-15}$ (Fig.~\ref{ExIsingW1} (C.I, C.II, and C.III)). 

Ideal Majorana behavior is realized only for perfectly localized left- or right-edge modes. As Figs.~\ref{ExIsingW1} and \ref{ExIsingW2} demonstrate, modes with small but non-zero energy may still exhibit edge localization but do not fully satisfy the Majorana condition. In such cases, the semi-infinite recursion relation does not strictly apply; instead, the modes satisfy the coupled finite-size equations, although their qualitative features can still be inferred directly from the Hamiltonian.

\vspace{0.2cm}

\noindent \textbf{Connection to the semi-infinite limit}

\begin{figure}[htbp]
	\centering
		\includegraphics[scale = 0.55]{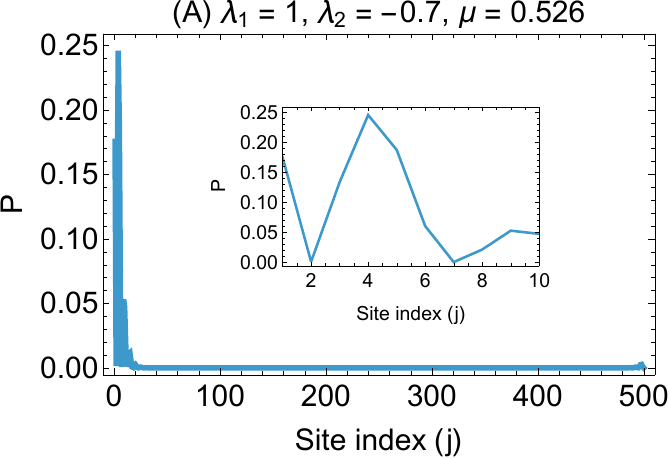}
		\hspace{1.2cm}
		\includegraphics[scale = 0.55]{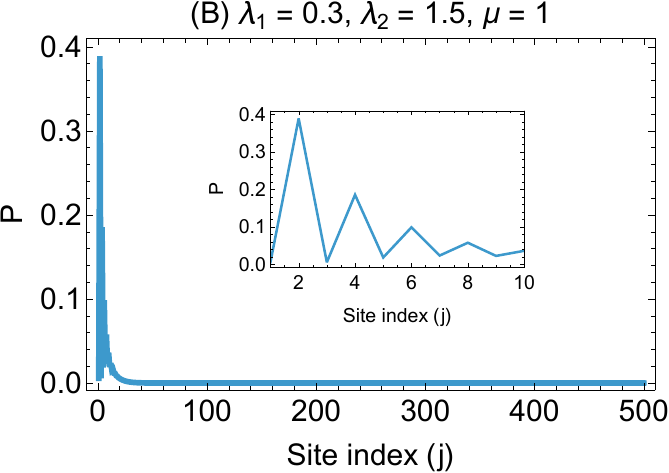}
	\caption{\justifying{Two eigenmodes corresponding to the parameter values in Fig.~\ref{recursionG1} and Fig.~\ref{recursionG3}, plotted for a finite system size. The corresponding eigen energies are $-3.01891 \times 10^{-16}$ and $-1.33435 \times 10^{-15}$, respectively. The inset shows a magnified view highlighting the close agreement between the finite-size numerical eigenmodes and the zero-mode profiles obtained from the recursion relations in Fig.~\ref{recursionG1} and Fig.~\ref{recursionG3}. The small discrepancy in the lattice sites at which the probability maxima occur arises because, in the semi-infinite recursion approach, the relative amplitudes of $A_1$ and $A_2$ can be chosen freely, whereas in a finite chain they are constrained by the global eigenvalue problem. Despite this difference, both approaches consistently demonstrate an intrinsic parameter-controlled peak shift.}}
	\label{realspacecomp}
\end{figure}

\begin{figure}[htbp]
	\centering
		\includegraphics[scale = 0.55]{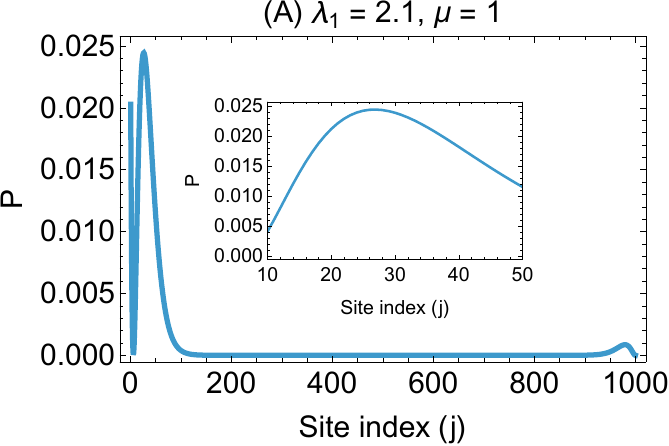}
		\hspace{1.2cm}
		\includegraphics[scale = 0.55]{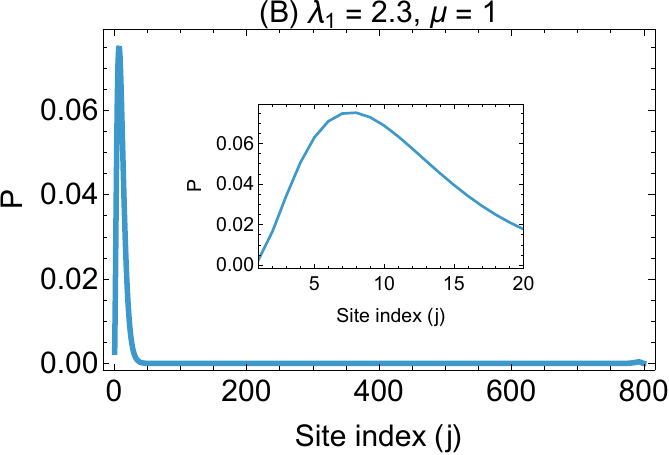}
	\caption{\justifying{Two eigenmodes for parameter values corresponding to Fig.~\ref{recursiondashed}, lying on the dashed curve $\lambda_2 = -\lambda_1^2/4\mu$ in Fig.~\ref{phase}(A), plotted for a finite chain. The corresponding eigen energies are $-2.21723 \times 10^{-16}$ and $-1.88176 \times 10^{-16}$, respectively. The profiles show a pronounced shift of the probability maxima away from the chain ends and into the lattice interior, while the modes retain an exponentially decaying envelope characteristic of boundary-origin zero modes. The spatial features are highly sensitive to parameter variations and persist only when the system lies on, or is sufficiently close to, the dashed curve.}}
	\label{realspacecurve}
\end{figure}

Once the system size exceeds a parameter-dependent crossover length scale determined by the characteristic roots, the overlapping finite-size modes approach well-defined left- or right-localized modes. These modes recover all intrinsic spatial features predicted analytically in the semi-infinite limit, including oscillatory decay, monotonic decay, perfect localization, and inward-shifted probability maxima (Figs.~\ref{realspacecomp} and \ref{realspacecurve}). The nature of the modes may also evolve with system size; for example, a mode can transition between oscillatory decay and monotonic decay. 

\section{Discussion}
\label{discussion}
The winding number, which determines the number of boundary MZMs, remains constant within a given topological phase and changes only at phase transitions. However, the winding number alone does not fix the spatial structure of the corresponding MZMs. Even within a single topological phase, the profiles of MZMS can vary continuously as system parameters are changed, without altering the topological classification. This variability originates from continuous changes in the roots of the characteristic equation associated with the MZM recursion relation. 

\subsection{Intrinsic Majorana zero-mode structure from the semi-infinite analysis}
The semi-infinite chain provides a natural framework for revealing the intrinsic spatial structure of zero modes, determined solely by the Hamiltonian. In this limit, boundary-induced hybridization between left- and right-origin modes is absent, allowing exact zero-energy solutions whenever the recursion admits normalizable modes. The characteristic roots directly encode the decay length, oscillatory behavior, and interference structure of these zero modes. Continuous changes in the magnitude or phase of the roots deform the spatial profile smoothly without affecting their topological origin. Importantly, this continuous variation of the roots also provides quantitative information about proximity to phase transitions. As the roots approach the unit circle, the decay length diverges, signaling criticality. Away from criticality, the roots can define the length scales that determine the realistic system size at which to observe these phenomena in realistic chains. 

Perfectly localized MZMs arise as special solutions of the recursion relation when destructive interference truncates the wavefunctions after a finite number of sites. In the extended Kitaev chain considered here, such perfectly localized MZMs occur over a finite region of parameter space, in contrast to the nearest-neighbor Kitaev model, where they appear only at a single fine-tuned point corresponding to zero chemical potential. In phases supporting two MZMs, one or both MZMs may be perfectly localized depending on the parameters and on the choice of relative amplitudes at the first two lattice sites. These intrinsic features form the basis for understanding the more complex spatial structures that arise in finite systems, discussed in the following subsections.

Beyond perfectly localized solutions, the semi-infinite analysis reveals a more subtle intrinsic feature: even when MZMs are normalizable and well separated, their probability maxima need not be located at the chain ends. In particular, boundary-origin MZMs can exhibit probability maxima at interior lattice sites. This behavior arises from interference between multiple admissible solutions of the recursion relation, which can relocate the position of maximum probability away from the boundary. This shift can be tuned continuously by varying system parameters and, in certain regimes, may place the peak deep inside the lattice.

Crucially, this behavior does not indicate the presence of true bulk-localized Majorana modes. The system remains completely uniform, and the modes retain an exponentially decaying envelope away from the peak, as required by bulk-boundary correspondence. The shifted peak reflects the internal structure of boundary-origin MZMs, rather than localization induced by defects, vortices, or junctions, in which spatial pinning arises from explicit inhomogeneities or topological defects in the Hamiltonian. These shifted MZMs, therefore, remain boundary-origin MZMs, but with spatial profiles that are not accessible in the conventional nearest-neighbor Kitaev chain.

\subsection{Implications for numerical and experimental observations}

The finite-size effects, discussed in Section~\ref{sizemode}, imply that numerical and experimental observations may not directly reflect the intrinsic MZM structure of the underlying Hamiltonian. In particular, boundary-induced hybridization can mask features such as perfect localization or inward-shifted probability maxima when the system size is not large compared to the characteristic decay length.

The analytical solutions developed here provide a complementary perspective by identifying the characteristic roots and the associated length scales that govern the intrinsic structure of MZMs. This enables a clear distinction between genuine Hamiltonian-controlled features and those arising from finite-size effects, and provides a reliable framework for interpreting numerical simulations and experimental observations of MZMs.

\subsection{Robustness against weak disorder}
We briefly comment on the effect of weak disorder, which is unavoidable in realistic implementations of one-dimensional topological superconductors. It is well established that MZMs in systems belonging to class BDI remain robust against weak disorder, provided the bulk energy gap remains open, and the protecting symmetries are preserved~\cite{PhysRevB.93.075129, PhysRevB.96.241113, PhysRevB.100.205302, Levy2019, PhysRevB.103.224505}. In particular, a disorder that preserves the protecting symmetries does not alter the topological invariant and hence does not remove the boundary modes~\cite{Brouwer2011PRL}. In the present framework, the spatial structure of the zero modes is governed by the characteristic roots of the recursion relation. Weak disorder can be expected to introduce site-dependent perturbations to the effective parameters entering the recursion relation, thereby locally modifying these roots. This leads to quantitative changes in the localization length and oscillatory behavior of the modes. However, the features identified in this work, namely, the existence of boundary-origin MZMs and the possibility of inward-shifted probability maxima, do not rely on fine-tuned parameter choices. These features are therefore expected to persist under weak disorder at a qualitative level, although their detailed spatial profiles may be modified. A systematic analysis of disorder effects within the recursion-relation framework, particularly for extended Kitaev chains with longer-range couplings, is left for future work.

\subsection{Comparison with previous works}

Ref.~\cite{mahyaeh2018zero} presents exact solutions for finite chains with next-nearest-neighbor interactions. However, because the analysis is performed in a finite-size setting, the solutions are not expressed in closed form, which precludes a direct analytical exploration of the full parameter space. As a result, the spatial structure of the modes is examined at selected parameter values, and a systematic scan of the phase diagram is not feasible. Consequently, features such as inward-shifted probability maxima, which require identifying specific regions in parameter space, are not explicitly observed. In contrast, the semi-infinite recursion framework developed here provides direct analytical control over the characteristic roots, enabling a complete and systematic classification of spatial profiles across the phase diagram. Importantly, these features are also recovered in finite chains once the system size exceeds a parameter-dependent crossover length scale, demonstrating that they are not merely idealized properties of the semi-infinite limit. 

In Ref.~\cite{leumer2020exact}, exact modes were studied for a finite Kitaev chain with unequal pairing and hopping amplitudes. Intrinsic peak shifts were not observed, likely due to the absence of next-nearest-neighbor interactions. Moreover, the site-dependent oscillatory terms in their analytical expressions (Eqs.~112 and 113) are independent of system parameters, limiting the range of accessible spatial profiles. Ref.~\cite{Alecce_2017} considers long-range ($r$-neighbor) interactions under specific constraints on coupling strengths in a finite chain. In this case, the recursion relation reduces to a higher-order polynomial, making a general analytical treatment difficult.

\section{Summary and Conclusion}
In summary, zero modes, edge modes, and Majorana zero modes need not coincide in their spatial characteristics, even within a fixed topological phase in finite chains. To understand the cause of their difference, an intrinsic study is required, which is carried out by solving the recursion relation in the semi-infinite chain limit. While symmetry determines the topological invariants of the phase, which in turn determine the number of MZMs, their spatial structure is determined by the Hamiltonian parameters. As a result, MZMs within the same phase can display qualitatively different behaviors, including monotonic decay, oscillatory decay, and perfect localization.
A key outcome of this work is that boundary-origin MZMs need not have their maximum probability at the chain edge. Instead, the maximum can occur at interior lattice sites, with the probability decaying away from its peak. This shows that identifying edge modes solely by a maximum at the first site can be misleading, even in completely uniform systems.
Importantly, the analytical framework developed here provides a direct link between the intrinsic MZM structure obtained in the semi-infinite limit and the behavior observed in finite systems. It clarifies how features such as shifted probability maxima or extended profiles can arise intrinsically from the Hamiltonian, and how finite-size effects influence their observation.
In simple terms, the spatial structure of MZMs is determined by the combination and interference of multiple decay components, which can produce a range of distinct profiles even without changing the topological phase.

\section*{Acknowledgments}

VP would like to thank Professor R. Srikanth for financial support through the Indian Science and Engineering Research Board (SERB) grant CRG/2022/008345. SS would like to thank CRG/2021/000996. 

\begin{comment}
\section*{Data Availability}
All data generated or analyzed during this study are included in this article. The results are based on analytical calculations and numerical simulations implemented in Mathematica, and the corresponding codes can be provided upon reasonable request.

\section*{Author contributions}
V.P and S.S. identified the problem. V.P., V.M., and S.S solved the problem. V.P and V.M generated all the figures and wrote the manuscript. 

\end{comment}

\appendix

\section{Critical lines and points}\label{critical lines}
The phase transition points are determined by the closing of the bulk energy gap, which occurs when the excitation spectrum vanishes. From Eq.~\ref{spectra}, this requires both $\chi_y(k)$ and $\chi_z(k)$ to be zero simultaneously. The condition $\chi_y(k)=0$ is satisfied for $k=0, \pm \pi$ and $k=\cos^{-1}\bigg(\dfrac{-\lambda_1}{2\lambda_2}\bigg)$. The corresponding phase transition lines are obtained by imposing $\chi_z(k)$ at these values:
\begin{itemize}
    \item For $k=0$, $\chi_z(k) = -2\lambda_1 - 2\lambda_2 + 2\mu$. The gap closes when $\lambda_1 + \lambda_2 = \mu$, corresponding to the line $C_2$ in Fig.~\ref{phase}.

    \item For $k=\pm \pi$, $\chi_z(k) = 2\lambda_1 - 2\lambda_2 + 2\mu$. The gap closes when $\lambda_2 - \lambda_1 = \mu$, corresponding to the line $C_1$ in Fig.~\ref{phase}.

    \item For $k=\cos^{-1}\bigg(\dfrac{-\lambda_1}{2\lambda_2}\bigg)$, $\chi_z(k)=2\lambda_2 + 2\mu$. The gap closes when $\lambda_2 + \mu = 0$, corresponding to the line $C_3$ in Fig.~\ref{phase}, with the constraint $\lambda_1\leq 2\lambda_2$ to ensure a real solution for $k$.

    \item The multicritical points correspond to intersections of these lines. The point $M_1$ arises when the gap closes simultaneously at $k=0$ and $\pm \pi$, whereas $M_2$ corresponds to the condition where the $C_3$ line intersects with the $k=0$ gap-closing condition. See Fig.~\ref{phase}
\end{itemize}

\section{Majorana basis transformation}\label{MajoranaBasis}
In this appendix, we rewrite the fermionic Hamiltonian Eq.~\ref{Eq.Hamiltonian_Fermion_Basis} in the main text (expanded form in Eq.~\ref{appendix_H_fermionic}) in terms of Majorana operators,

\begin{equation}\label{appendix_H_fermionic}
\begin{split}
    H=&-\mu \sum_{j=1}^N (1-2c_j^{\dagger}c_j)-\lambda_1\sum_{j=1}^{N-1}(c_{j}^{\dagger}c_{j+1} + c_{j}^{\dagger}c_{j+1}^{\dagger} + c_{j+1}^{\dagger}c_{j} + c_{j+1}c_{j})\\
    &- \lambda_2\sum_{j=2}^{N-1}(c_{j-1}^{\dagger}c_{j+1} + c_{j+1}^{\dagger}c_{j-1}^{\dagger} + c_{j+1}^{\dagger}c_{j-1} + c_{j-1}c_{j+1})
\end{split}
\end{equation}

We define the normalized operators
 \[
 a_i = \dfrac{c_i + c_i^{\dagger}}{\sqrt{2}}, \quad 
 b_i= \dfrac{-i(c_i^{\dagger} - c_i)}{\sqrt{2}},
 \]
which satisfy the anti-commutation relations
\[
    a_j^2 = b_j^2 = \frac{1}{2}, \quad \{a_i, b_j\} = 0, \quad
    \{a_i, a_j\} = \{b_i, b_j\} = \delta_{i, j}.
\]    

The inverse transformation is given by
\[
    c_i = \frac{1}{\sqrt{2}}(a_i - ib_i), \quad
    c_i^{\dagger} = \frac{1}{\sqrt{2}}(a_i + ib_i).
\]

Substituting these expressions into Eq.~\ref{appendix_H_fermionic}, we evaluate each term separately.

\vspace{0.2cm}

\noindent \textbf{Chemical potential term:}

\begin{equation*}
    H =  -\mu \sum_{j=1}^N (1-2c_j^{\dagger}c_j)
    =-\mu \sum_{j=1}^N (c_jc_j^{\dagger}-c_j^{\dagger}c_j)
\end{equation*}

Substituting the Majorana operators,
\begin{equation*}
  H=  -\frac{\mu}{2} \sum_{j=1}^N ((a_j-ib_j)(a_j+ib_j)-(a_j+ib_j)(a_j-ib_j))= -i\mu \sum_{j=1}^{N}(a_jb_j-b_j a_j) 
\end{equation*}
The constant term in the fermionic Hamiltonian is implicitly accounted for in this transformation and does not affect the eigenvalues.

\vspace{0.2cm}

\noindent \textbf{Hopping and pairing terms:}

For the remaining terms, substituting into Eq.~\ref{appendix_H_fermionic} and using the anti-commutation relations, terms of the form $a_ia_j$ and $b_ib_j$ cancel pairwise due to antisymmetry of the operators. The surviving contributions involve mixed products of $a$ and $b$ operators.

This yields
\begin{align*}
    \begin{split}
     H &= -i\mu \sum_{j=1}^{N}(a_jb_j-b_j a_j) - i\lambda_1\sum_{j = 1}^{N-1}(b_j a_{j+1} - a_{j+1}b_j) - i\lambda_2\sum_{j=2}^{N-1}( b_{j-1}a_{j+1} - a_{j+1}b_{j-1}) \\   
     H &= -2i\left[ -\mu\sum_{j=1}^N  b_j a_j +  \lambda_1 \sum_{j=1}^{N-1} b_j a_{j+1} +  \lambda_2\sum_{j=2}^{N-1}b_{j-1}a_{j+1}\right]
     \end{split}
\end{align*}

This is the Majorana representation of the Hamiltonian used in Sec.~\ref{majoranabases}.

\section{Recursion relation and its solution}\label{RecursionSolution}
In this appendix, we derive the recursion relations for the Majorana amplitudes starting from the Hamiltonian written in the Majorana basis (Eq.~\ref{FermionH} in the main text).

The Hamiltonian can be expressed in matrix form as
\begin{align*}
H = -i
\left(
\begin{array}{cccccc}
0 & \mu & 0 & \cdots &  &  \\
-\mu & 0 & \lambda_1 & 0 & \lambda_2 & \cdots \\
0 & -\lambda_1 & 0 & \mu & 0 & \cdots \\
0 & 0 & -\mu & 0 & \lambda_1 & 0 \\
\vdots &  & \vdots &  & \ddots & \ddots \\
\end{array}
\right).
\label{eq:matrix17}
\end{align*}
We consider a general eigenvector of the form $\psi=(A_1, A_1^{'}, A_2, A_2^{'}, \cdot\cdot\cdot, A_N, A_N^{'})^T$. The eigenvalue equation $H\psi=E\psi$ then leads to a set of coupled equations for the amplitudes. By examining the structure of the matrix, one finds that the equations obtained from the odd and even rows decouple into two recursion relations:
\begin{subequations}
    \begin{align}
    -\mu A_j + \lambda_1A_{j+1} + \lambda_2A_{j+2} = & iE A_{j}^{'} \\
    \mu A_{j}^{'} - \lambda_1A_{j-1}^{'} - \lambda_2A_{j-2}^{'} = &  iE A_{j}
    \end{align}
\end{subequations}
These equations correspond to Eq.~\ref{Majoranaeigenvector} in the main text.

The boundary conditions arise from the chain's finite size. For $j=N$, the first equation formally involves amplitudes $A_{N+1}$ and $A_{N+2}$, which lie outside the chain. These must therefore vanish, i.e., $A_{N+1}=0=A_{N+2}$. Similarly, for the $j=1$, the second equation involves $A_{0}$ and $A_{-1}$, which are not physical sites, and hence $A_{-1}= 0= A_{0}$.

\section{Number of zero modes}\label{w}
To determine the number of zero-energy modes, we analyze the normalizability of the solutions obtained from the characteristic roots $q_\pm$. We assume $\lambda_1\geq 0$ and $\mu\geq0$.
\begin{itemize}
    \item \textbf{Region $G_1$: } $\lambda_2 > \mu + \lambda_1$. In this case, both $|q_{\pm}|<1$, and hence there are two normalizable modes.

    \item \textbf{Line $C_1$: } $\lambda_2 = \mu + \lambda_1$. Here, $q_+ = \dfrac{\mu}{\mu + \lambda_1}$ and $q_- = -1$. Only one root satisfies $|q| < 1$, so there is one normalizable mode.

    \item \textbf{Region $G_4$: } $\mu - \lambda_1 < \lambda_2 < \mu + \lambda_1$. In this region, $|q_+| < 1$ while $|q_-| > 1$, so there is one normalizable mode.

    \item \textbf{Point $M_1$: } $\lambda_1 = 0$ and $\lambda_2 = \mu$. Here, $q_\pm = \dfrac{|\mu|}{\mu}$, so, $|q_\pm| = 1$. Therefore, there are no normalizable modes.

    \item \textbf{Region $G_2$: } $-\mu < \lambda_2 < \mu-\lambda_1$. In this case, both $|q_\pm| > 1$, and hence there are no normalizable modes.

    \item \textbf{Line $C_2$: } $\lambda_2 = \mu - \lambda_1$ with $\lambda_1 < 2\mu$. Here, $|q_+| = 1$ and $|q_-| > 1$, so there are no normalizable modes. 
    
    \item \textbf{Line $C_4$: } $\lambda_2 = \mu - \lambda_1$ with $\lambda_1 > 2 \mu$. In this case, $|q_+| < 1$ and $|q_-| = 1$, so there is one normalizable mode. 
    
    \item \textbf{Point $M_2$: } $\lambda_2 = \mu - \lambda_1$ and $\lambda_1 = 2 \mu$. Both $|q_\pm| = 1$, so there are no normalizable modes. 

    \item \textbf{Line $C_3$: } $\lambda_2 = -\mu$. There are two cases: (a) If $\lambda_1 < 2\mu$, the roots become complex with $|q_\pm| = 1$, and hence there are no normalizable modes; 
    (b) If $\lambda_1 > 2\mu$, the roots remain real with $|q_+| < 1$ and $|q_-| > 1$, so there is one normalizable mode.

    \item \textbf{Region $G_3$: } $\lambda_2 < \mu - \lambda_1$ and $\lambda_2 < -\mu$. In this region, both $|q_\pm| < 1$, hence there are two normalizable modes. 

    \item \textbf{Degeneracy line: } The dashed curve in Fig.~\ref{phase}(A) corresponds to the condition $q_+ = q_-$. Below this curve, the roots are complex conjugate with $|q_{+}|=|q_{-}|$, while above it, the roots are real.
\end{itemize}

\section{\texorpdfstring{General solution for $\text{w}=2$}{}}\label{w2generalsolution}
In the $\text{w}=2$ regions, the recursion relation admits two normalizable roots. The general solution for the amplitude at the $j$-th site can therefore be written as
\begin{equation*}
    A_j=c_1 q_+^j + c_2 q_-^j
\end{equation*}

The coefficients can be expressed in terms of the physically relevant amplitudes at the first two sites
\[
 A_1 = c_1 q_+ + c_2 q _-, \quad 
 A_2 = c_1 q_+^2 + c_2 q_-^2.
 \]
Solving for $c_1$ and $c_2$, 
\[
c_1 = \frac{A_1 q_- - A_2}{q_+(q_- - q_+)}, \quad 
c_2 = \frac{A_1 q_+ - A_2}{q_-(q_+ - q_-)}.
\]
Substituting back, the solution becomes
\begin{equation*}
    A_j = \frac{1}{q_+ - q_-}\big[A_1(q_+ q_-^{j-1} - q_-q_+^{j-1}) + A_2(q_+^{j-1} - q_-^{j-1})\big]
\end{equation*}
provided $q_+ \neq q_-$. This solution can be further simplified depending on the parameter regime.

\subsection{\texorpdfstring{Complex roots: $\lambda_1^2 + 4\mu \lambda_2 < 0$}{Complex roots}}\label{w2complexq}

Since $\mu > 0$, this condition requires $\lambda_2<0$. The roots are complex and can be written as $q_\pm = R e^{\pm i\theta}$, where 
\[
R = \sqrt{\frac{-\mu}{\lambda_2}}, \quad 
\cos{\theta} = \frac{\lambda_1}{\sqrt{-4\mu\lambda_2}}.
\]

The terms in the general solution become
\[
(q_+q_-^{j-1} - q_-q_+^{j-1}) = 2i R^j \sin{(j-2)\theta}, 
\]
\[
(q_+^{j-1} - q_-^{j-1}) = 2i R^{j-1} \sin{(j-1)\theta}, 
\]
\[ 
q_+ - q_- = 2i R \sin{\theta}. 
\]

Using the identity 
\[
\sin{(j-2)\theta} = \sin{(j-1)\theta}\cos{\theta} - \cos{(j-1)\theta}\sin{\theta}, 
\]
the solution reduces to 
\[
A_j = \frac{R^{j-2}}{\sin{\theta}}\big[(A_1 R \cos{\theta} + A_2)\sin{(j-1)\theta} - A_1 R \sin{\theta}\cos{(j-1)\theta}\big].
\]

Defining 
\[
y = \sqrt{(A_1 R \cos{\theta} + A_2)^2 + (A_1 R \sin{\theta})^2}, \quad
\cos{\phi} = \frac{A_1 R \cos{\theta} + A_2}{y},
\]

We obtain
\begin{equation*}
    A_j = \frac{R^{j-2}}{\sin{\theta}} \, y \sin{[(j-1)\theta - \phi]}
\end{equation*}

\subsection{\texorpdfstring{Real root: $\lambda_1^2 + 4\mu \lambda_2 > 0$}{Real roots}}\label{w2realq}

In this case, both roots are real and satisfy $|q| < 1$. We use the identity
\[
q_+^{j} - q_-^{j} = (q_+ - q_-)q_+^{j-1} + q_-(q_+^{j-1} - q_-^{j-1}),
\]
to rewrite the general solution.

Starting from 
\[
A_j=\frac{1}{q_+ - q_-}\big[-A_1q_+q_-(q_+^{j-2} - q_-^{j-2}) + A_2(q_+^{j-1} - q_-^{j-1})\big],
\]
We apply the identity again to the terms $(q_+^{j-2} - q_-^{j-2})$ and $(q_+^{j-1} - q_-^{j-1})$, which gives
\begin{equation*}
    \begin{split}
    A_j = &\frac{1}{q_+ - q_-}\big[-A_1 q_+q_-\big((q_+-q_-)q_+^{j-3} + q_-(q_+^{j-3} - q_-^{j-3})\big) + A_2\big((q_+-q_-)q_+^{j-2} + q_-(q_+^{j-2} - q_-^{j-2})\big)\big]
    \end{split}
\end{equation*}

After rearranging terms, we obtain
\begin{equation*}
    A_j = \frac{q_-}{q_+ - q_-}\big[-A_1 q_+q_-(q_+^{j-3} - q_-^{j-3}) + A_2(q_+^{j-2}- q_-^{j-2})\big] + q_+^{j-2}\big[-A_1q_- + A_2\big]
\end{equation*}

This leads to the recursion relation
\[
A_j= q_-A_{j-1} + (A_2 - q_-A_1) q_+^{j-2}.
\]

The solution of this recurrence is 
\begin{equation*}
    A_j = q_-^{j-2} A_2 + \sum_{k=1}^{j-2} f(k) q_-^{j-k-2}
\end{equation*}
where, $f(k) = (A_2 - A_1q_-)q_+^{k}$

\subsection{\texorpdfstring{Degenerate roots: $\lambda_1^2 + 4\mu \lambda_2 = 0$ and $\lambda_2 < -\mu$}{Degenerate roots}}\label{w2equalq}

Along the dashed curve in Fig.~\ref{phase} (A), the roots become degenerate, $q_+ = q = q_-$, and the above-mentioned general solution is not applicable. The general solution in this case takes the form
\[
A_j=(c_1 + c_2j)q^j,
\]
since both $q^j$ and $jq^j$ satisfy the recursion relation.

Using
\[
A_1 = (c_1+c_2)q, \quad 
A_2 = (c_1 + 2c_2)q^2,
\]
We obtain
\[
c_1 = \frac{2 A_1 q - A_2}{q^2}, \quad 
c_2 = \frac{A_2 - A_1 q}{q^2}.
\]

Substituting back, 
\begin{equation*}
    A_j=\big[(2-j) q A_1 + (j-1) A_2 \big]q^{j-2}
\end{equation*}

\bibliographystyle{apsrev4-2}
\bibliography{references}

\end{document}